\documentclass{aa}

\usepackage{graphicx}
\usepackage{epsfig}
\usepackage{natbib}
\usepackage{txfonts}

\newcommand{\logg}{\ensuremath{\log g}}

\newcommand{\mlp}{\ensuremath{\alpha_{\mathrm{MLT}}}}

\newcommand{\moh}{\ensuremath{[\mathrm{M/H}]}}
\newcommand{\feoh}{\ensuremath{[\mathrm{Fe/H}]}}

\newcommand{\Teff}{\ensuremath{T_{\mathrm{eff}}}}
\newcommand{\tauross}{\ensuremath{\tau_{\mathrm{Ross}}}}

\newcommand{\beq}{\begin{equation}}
\newcommand{\eeq}{\end{equation}}

\newcommand{\xtmean}[1]{\ensuremath{\left\langle #1\right\rangle}}

\newcommand{\MoH}{\ensuremath{\left[\mathrm{M}/\mathrm{H}\right]}}

\newcommand{\tauc} {\tau_{\rm c}}
\newcommand{\taul} {\tau_{\ell}}
\newcommand{\uclam} {u_{\rm c}}
\newcommand{\diff}[1]{{\rm d}\, #1}

\newcommand{\COBOLD}{{\tt CO$^5$BOLD}}
\newcommand{\LHD}{{\tt LHD}}

\newcommand{\MARCS}{{\tt MARCS}}

\newcommand{\LINFOR}{{\tt Linfor3D}}

\begin{document}

\title{Three-dimensional hydrodynamical \COBOLD\ model atmospheres of red giant stars\\}

\subtitle{II. Spectral line formation in the atmosphere of a giant located near the RGB tip}

\author{A.~Ku\v{c}inskas\inst{1,2}
        \and
        M.~Steffen\inst{3}
        \and
        H.-G.~Ludwig\inst{4}
        \and
        V.~Dobrovolskas\inst{2}
        \and
        A.~Ivanauskas\inst{2,1}
        \and
        J.~Klevas\inst{1}
        \and
        D.~Prakapavi\v{c}ius\inst{1}
        \and\\
        E.~Caffau\inst{4}
        \and
        P.~Bonifacio\inst{5}
       }

\offprints{A. Ku\v{c}inskas}

\institute{
        Vilnius University Institute of Theoretical Physics and Astronomy, A. Go\v {s}tauto 12, Vilnius LT-01108, Lithuania\\
        \email{arunas.kucinskas,augustinas.ivanauskas,jonas.klevas,dainius.prakapavicius@tfai.vu.lt}
        \and
        Vilnius University Astronomical Observatory, M. K. \v{C}iurlionio 29, Vilnius LT-03100, Lithuania\\
        \email{Vidas.Dobrovolskas@ff.vu.lt}
        \and
        Leibniz-Institut f\"ur Astrophysik Potsdam, An der Sternwarte 16, D-14482 Potsdam, Germany\\
        \email{msteffen@aip.de}
        \and
        ZAH Landessternwarte K\"{o}nigstuhl, D-69117 Heidelberg, Germany\\
        \email{hludwig,elcaffau@lsw.uni-heidelberg.de}
        \and
        GEPI, Observatoire de Paris, CNRS, Universit\'{e} Paris Diderot, Place Jules Janssen, 92190 Meudon, France\\
        \email{piercarlo.bonifacio@obspm.fr}\\
        }

\date{Received: date; accepted: date}

\abstract
   {}
   {
   We investigate the role of convection in the formation of atomic and molecular
   lines in the atmosphere of a red giant star. For this purpose we study
   the formation properties of spectral lines that belong to a number of
   astrophysically important tracer elements, including neutral and singly
   ionized atoms (\ion{Li}{i}, \ion{N}{i}, \ion{O}{i}, \ion{Na}{i}, \ion{Mg}{i},
   \ion{Al}{i}, \ion{Si}{i}, \ion{Si}{ii}, \ion{S}{i}, \ion{K}{i}, \ion{Ca}{i},
   \ion{Ca}{ii}, \ion{Ti}{i}, \ion{Ti}{ii}, \ion{Cr}{i}, \ion{Cr}{ii},
   \ion{Mn}{i}, \ion{Fe}{i}, \ion{Fe}{ii}, \ion{Co}{i}, \ion{Ni}{i},
   \ion{Zn}{i}, \ion{Sr}{ii}, \ion{Ba}{ii}, and \ion{Eu}{ii}), and molecules
   (CH, CO, C$_2$, NH, CN, and OH).
   }
   {
   We focus our investigation on a prototypical red giant located close to the
   red giant branch (RGB) tip (\Teff=3660\,K, \logg=1.0, \moh=0.0).
   We used two types of model atmospheres, 3D hydrodynamical and classical 1D,
   calculated with the \COBOLD\ and \LHD\ stellar atmosphere codes, respectively.
   Both codes share the same atmospheric parameters, chemical composition, equation
   of state, and opacities, which allowed us to make a strictly differential
   comparison between the line formation properties predicted in 3D and 1D.
   The influence of convection on the spectral line formation was assessed with
   the aid of 3D--1D abundance corrections, which measure the difference between
   the abundances of chemical species derived with the 3D hydrodynamical and
   1D classical model atmospheres.
   }
   {
   We find that convection plays a significant role in the spectral line
   formation in this particular red giant. The derived 3D--1D abundance
   corrections rarely exceed $\pm0.1$\,dex when lines of neutral atoms and
   molecules are considered, which is in line with the previous findings for
   solar-metallicity red giants located on the lower RGB. The situation is
   different with lines that belong to ionized atoms, or to
   neutral atoms with high ionization potential. In both cases, the
   corrections for high-excitation lines ($\chi > 8$~eV) may amount to
   $\Delta_{\rm 3D-1D}\sim-0.4$\,dex. The 3D--1D abundance corrections
   generally show a significant wavelength dependence; in most cases they
   are smaller in the near-infrared, at 1600--2500~nm.
   }
   {}


   \maketitle

\section{Introduction}

Current three-dimensional (3D) hydrodynamical codes have brought a new level
of realism to the modeling of stellar atmospheres, making it possible to
assess the influence of various nonstationary phenomena on the observable
properties of various classes of stars. Since red giants are amongst the
brightest objects in intermediate age and old populations, precise
understanding of their internal structures and observable properties is of
key importance for studies of stellar populations in the Galaxy and beyond.
Three-dimensional hydrodynamical stellar model atmospheres may prove especially
useful in this context, as they can provide important theoretical insights
about how and to which extent convection and other nonstationary phenomena
may influence the observable properties of red giant stars.

Up to now, only a few studies have focused on the investigation of observable
properties of red giants with 3D hydrodynamical stellar model atmospheres.
In one of the early attempts, \citet[]{KHL05} have found that convection may
noticeably alter the thermal structure of the red giant atmosphere which, in
turn, may affect the spectral energy distribution of the emerging radiation
field. This may yield a difference of $\sim0.2-0.3$\,mag in broad-band
photometric colors predicted by the 3D hydrodynamical and classical
1D models. It was later shown by \citet[]{CAT07} that convection in the
atmospheres of somewhat warmer red giants ($\Teff\approx4700-5100$\,K,
$\logg=2.2$, $\feoh=0.0$ to $-3.0$) may lead to significant changes in their
thermal structures, especially at lowest metallicities. Remarkably, at
$\moh=-3.0$ this may result in differences of up to $-1.0$\,dex in the
abundances of chemical elements derived with the 3D hydrodynamical and
classical 1D model atmospheres. This result was later confirmed by
\citet[]{DKL10} and \citet[]{IKL10}, who used different 3D hydrodynamical
stellar atmosphere and spectrum synthesis codes, \COBOLD\ and \LINFOR,
respectively, together with the atmospheric parameters of red giants similar
to those utilized by \citet[]{CAT07}. All these studies unequivocally point
to the fact that the role of convection in the atmospheres of red giant stars
may be significantly more important than previously thought, especially at
metallicities $\moh\lesssim-1.0$.

Aiming at a more systematic study, we have recently started a project to
investigate internal structures and spectral properties of red giants
across the HR diagram with the aid of 3D hydrodynamical model
atmospheres. For this purpose we used the 3D hydrodynamical stellar
atmosphere code \COBOLD\ to produce a grid of red giant model atmospheres,
which are now available as part of the CIFIST grid of \COBOLD\ 3D model
atmospheres \citep[][the grid is constantly updated with new
  models]{LCS09}. In the first study of this series \citep[]{LK12} we
investigated physical properties of convection in the atmosphere of a red
giant located close to the red giant branch (RGB) tip 
($T_{\rm eff}\approx3660$\,K, $\log g=1.0$, $\MoH=0$). In the present study
we extend our work on the same red giant and focus on the influence of
convection on the formation of various atomic and molecular lines. Our main
goal is to make a detailed comparison between the line strengths predicted by
the 3D hydrodynamical and classical 1D models, and discuss the consequences
for stellar abundance studies. General properties of the spectral energy
distribution and photometric colors of this particular red giant will be
discussed in a companion paper (Ku\v{c}inskas et al. 2012, in preparation).

The paper is organized as follows. The model atmospheres used in this work
are described in Sect.\,\ref{sect:procedure}, where we also outline the details
of spectrum synthesis calculations and the concept of 3D abundance corrections.
The main results are presented in Sect.\,\ref{sect:results}, including a brief
discussion of the basic properties of spectral line formation in the presence
of convection, and the differences between the predictions of the 3D
hydrodynamical and classical 1D models. The conclusions are presented in
Sect.\,\ref{sect:conclusions}. A more detailed analysis of the abundance
corrections derived in this work as well as additional background information
is given in Appendices~A, B, and C.

\section{Stellar model atmospheres and spectral line synthesis\label{sect:procedure}}

\subsection{Model atmospheres\label{sect:models}}

We used red giant model atmospheres calculated with the 3D
hydrodynamical \COBOLD\ and 1D stationary \LHD\ model atmosphere codes. Both
models were computed for the same atmospheric parameters (\Teff=3660\,K,
\logg=1.0, \moh=0.0) and elemental composition, and with the same equation of
state and opacities; they also share the same numerical radiative transfer
scheme. Additionally, we also used an average $\xtmean{\mbox{3D}}$ model,
which was produced by spatially averaging the 3D model structures at constant
Rosseland optical depth. The $\xtmean{\mbox{3D}}$ model does not contain
explicit information about the horizontal fluctuations of thermodynamical
and hydrodynamical quantities (e.g.\,temperature, pressure, velocity) that
are present in the full 3D model atmosphere. Therefore, the comparison of
the predictions of the full 3D and average $\xtmean{\mbox{3D}}$ models
allows one to estimate the relative importance of the horizontal temperature
fluctuations on the spectral line formation \citep[see, e.g.][for a more
detailed discussion]{CLS11}. Since the three types of models
used in this work are described in detail by \citet[][]{LK12}, we only
briefly summarize the most essential aspects of their calculation in the
following subsections.

\subsubsection{Three-dimensional hydrodynamical model and 3D snapshot
selection\label{sect:3dmodel}}

The red giant model atmosphere used in this work was calculated with the 3D
radiation hydrodynamics code \COBOLD\ \citep[][]{FSL12}. The \COBOLD\ code
uses a Riemann solver of Roe type to calculate the time evolution of the
hydrodynamical flow and the radiation field on a 3D Cartesian grid. The model
was computed using a grid of $150\times150\times151$ mesh points ($x\times
y\times z$), which corresponds to a physical box size of
$15.6\times15.6\times8.6$\,Gm$^3$. The radiative transfer is based on
monochromatic opacities from the \MARCS\ stellar atmosphere package
\citep{GEK08} which, to speed up the calculations, were grouped
into five opacity bins \citep[for details on the opacity binning approach see, e.g.]
[]{N82,L92,LJS94,VBS04}. Solar elemental abundances as given by
\citet{GS98} were assumed, except for carbon, nitrogen, and oxygen, for which
the following values were used: A(C)=8.41, A(N)=7.8, and A(O)=8.67 \citep{CLS08}.
The model calculations were made assuming local thermodynamic
equilibrium, LTE \citep[for more details about the model setup see][]{LK12}.

After the initial relaxation to a quasi-stationary state, we ran the model
simulations to cover a span of $\sim$$6\times10^6$\,sec ($\sim$70\,days)
in stellar time. This corresponds to approximately seven convective turnover
times\footnote{as measured by the Brunt-Vais\"{a}l\"{a} and/or advection
timescales, the latter equal to the time needed for the convective material
to cross 1.8 pressure scale heights \citep[see][]{LK12}.} in the
atmosphere of this particular red giant. From this sequence of relaxed
models we selected 14 3D model structures computed at different
instances in time (snapshots). Individual snapshots of this 14-snapshot
subset were then  used to produce average $\xtmean{\mbox{3D}}$
model and spectrum synthesis calculations (Sect.\,\ref{sect:<3d>model} and
\ref{sect:spectr-synth}, respectively). Snapshots of this 14-snapshot
ensemble are separated by $5\times10^5$\,sec ($\sim$6\,days) in stellar
time, which allows one to assume that in this subset they are statistically
uncorrelated. The snapshots were selected in such a way that most important
statistical properties of the snapshot ensemble, such as the average
effective temperature and its standard deviation, mean velocity at optical
depth unity, mean velocity profile and residual mass flux profile, would
match those of the entire 3D model run as closely as possible.

\subsubsection{Average $\xtmean{\mbox{3D}}$ model\label{sect:<3d>model}}

For each of the selected snapshots, the corresponding average 3D model,
$\xtmean{\mbox{3D}}$, was obtained by spatially averaging the thermal
structure of the 3D model box on surfaces of equal Rosseland optical
depth. We averaged the fourth moment of temperature and first moment
of gas pressure to preserve the radiative properties of the original
3D model, according to the prescription given in \citet{SLF95}. The
product of this procedure is a 1D model atmosphere that
retains the averaged vertical profile of the thermodynamical structure
of the 3D model, but lacks explicit information about the horizontal
inhomogeneities. This was done for the 14 3D model snapshots selected
at different instants in time (Sect.\,\ref{sect:3dmodel}), thus
obtaining a sequence of 14 $\xtmean{\mbox{3D}}$ models.

\subsubsection{One-dimensional (1D) model\label{sect:1dmodel}}

The 1D hydrostatical model was calculated with the \LHD\ code, using
the same atmospheric parameters, elemental abundances, opacities, and
equation of state as in the 3D model calculations described above
\citep[see, e.g.,] [for more details on the \LHD\ code]{CLS08}.
Convection in the \LHD\ models was treated according to the \citet{M78}
formulation of the mixing-length theory. The \LHD\ models were
calculated for the mixing-length parameters $\mlp=1.0$ and 2.0, to
investigate the influence of \mlp\ on the properties of line formation
in the 1D models (see Sect.\,\ref{sect:alphamlt}).

\subsection{Three- and one-dimensional spectrum synthesis
calculations\label{sect:spectr-synth}}

\subsubsection{Chemical elements and spectral line parameters}

We used fictitious spectral lines, i.e., lines of
a particular chemical element (molecule) for which the central
wavelength, $\lambda_{\rm c}$, the excitation potential of the
lower level, $\chi$, and line equivalent width, $W$, were selected
arbitrarily. This approach allowed us to cover a range in $\lambda_{\rm c}$
and $\chi$ to quantify the trends of line formation
properties in the 3D hydrodynamical and 1D classical model
atmospheres with respect to these two line parameters. Conceptually,
this method can be traced back to the work of \citet{SH02} and was
applied in several later studies too \citep[e.g.,][]{CAT07,DKL10,IKL10}.

Synthetic line profiles were calculated for the following tracer species of
astrophysical interest:
\begin{itemize}
\item
neutral atoms: \ion{Li}{i}, \ion{N}{i}, \ion{O}{i}, \ion{Na}{i}, \ion{Mg}{i},
\ion{Al}{i}, \ion{Si}{i}, \ion{S}{i}, \ion{K}{i}, \ion{Ca}{i}, \ion{Ti}{i},
\ion{Cr}{i}, \ion{Mn}{i}, \ion{Fe}{i}, \ion{Co}{i}, \ion{Ni}{i}, and \ion{Zn}{i};
\item
ionized atoms:
\ion{Si}{ii}, \ion{Ca}{ii}, \ion{Ti}{ii}, \ion{Cr}{ii}, \ion{Fe}{ii},
\ion{Sr}{ii}, \ion{Ba}{ii}, and \ion{Eu}{ii};
\item
molecules: CH, CO, C$_2$, NH, CN, and OH.
\end{itemize}

Fictitious lines were calculated at three wavelengths, $\lambda_{\rm c}=400$,
850, and 1600~nm. The first two  were selected to bracket the
wavelength range accessible with modern high-resolution optical spectrographs
(e.g., UVES/GIRAFFE@VLT, HARPS@ESO3.6m, HIRES@Keck), whereas the third
corresponds to the wavelength reachable with similar instruments in the
near-infrared $H$-band (e.g., CRIRES@VLT, NIRSPEC@Keck). In the case of
molecules, however, we used the wavelengths of real molecular bands in
the blue part of the spectrum, whenever available: the A--X OH transition
at 315\,nm \citep[e.g.,][]{BCG04}, the A--X transition of NH at 336\,nm
\citep[e.g.,][]{SCP05}, the B--X transition of CN at 388\,nm \citep{L68},
and the A--X electronic transition of CH at 432\,nm \citep{G85}. The
exceptions were C$_{\rm 2}$ and C, which do not have bands in the UV, and
for which we thus used the 400\,nm reference wavelength instead. One should
also note that 850 and 1600\,nm coincide respectively with the maximum
and minimum absorbtion of the H$^-$ ion, which is the most important
contributor to the continuum opacity in red giant atmospheres in the
optical to near-infrared wavelength range. On the other hand, the
continuum opacity at 400\,nm is dominated by the contribution from
metals. Therefore, the choice of the three wavelengths offers a
possibility to study the interplay between the different sources of
continuum opacity and line formation.

The excitation potentials were selected to cover the range of $\chi$
combinations possible for real (i.e., nonfictitious) elements/lines.
We used $\chi=0-6$\,eV for the neutral atoms, $0-10$\,eV for the ions
(in both cases with a step of $\Delta \chi=2$\,eV), and $0-4$\,eV
($\Delta \chi=1$\,eV) for the molecules. Exceptions were \ion{O}{i}
and \ion{N}{i}: their real lines are characterized by very high
excitation potentials ($\chi>9$\,eV) and thus the $\chi$ range for
these two species was chosen to be identical to that of the ions.

We finally stress that some of the elements/species studied here have
only a few real lines that can be used in the spectroscopic diagnostics
of red giant stars (e.g., \ion{Li}{i}, \ion{N}{i}, \ion{O}{i}, \ion{S}{i}).
Moreover, we also included several elements whose spectral lines are
inaccessible for observations in real life (e.g., \ion{K}{ii},
\ion{S}{ii}). We stress that all these elements were included to help
understand the trends and properties of spectral line formation in the
presence of realistically modeled convection, and to identify the
physical causes behind them. Obviously, the exact values of abundance
corrections for the combinations of atomic parameters where
spectral lines of these elements do not exist (or may not be observed
in stellar spectra) can only be of academical interest.

\subsubsection{Spectral line synthesis with 3D, $\xtmean{\mbox{3D}}$, and
1D model atmospheres}

Three-dimensional spectrum synthesis calculations are very time-consuming
and thus performing them by using the entire 3D model sequence would be
impractical. To make the task manageable, synthetic spectral line profiles
were computed using 14 3D model structures (snapshots) selected from
the sequence of seventy 3D models, fully relaxed to a quasi-stationary state
(see Sect.\,\ref{sect:3dmodel}). To speed-up calculations further, spectral
synthesis was carried out on a coarser $x,y$ grid of $50\times50$ points,
i.e. using only one third of the grid points of the original 3D model box
in each horizontal direction. Full-resolution test calculations performed
on the original $150\times150$ grid show that the reduced horizontal
resolution has a negligible effect on the properties of synthesized
spectral lines.

For all elements investigated here, spectral line profiles corresponding to
the 3D hydrodynamical model were calculated for each individual 3D model
structure (snapshot) in the 14-snapshot ensemble. A composite 3D line
profile was then constructed by co-adding the line profiles corresponding
to all fourteen snapshots. Similarly, $\xtmean{\mbox{3D}}$ line profiles
were calculated using a sequence of $\xtmean{\mbox{3D}}$ models obtained
according to the prescription given in Sect.\,\ref{sect:<3d>model}. Spectral
line profiles were calculated for each of the fourteen $\xtmean{\mbox{3D}}$
models and then co-added to produce a composite $\xtmean{\mbox{3D}}$ line
profile. A microturbulence velocity of $\xi_{\rm mic}=2.0$\,km/s was used
in the line synthesis calculations with $\xtmean{\mbox{3D}}$ and 1D models.
The choice of $\xi_{\rm mic}$ is not critical since we used only weak
unsaturated lines in the following (Sect.~\ref{sect:abund-corr-defin}).

To enable a strictly differential comparison between the predictions
of the 3D and 1D models, 3D, $\xtmean{\mbox{3D}}$, and 1D spectrum
synthesis computations were made with the same spectrum synthesis code, \LINFOR\footnote{http://www.aip.de/$\sim$mst/Linfor3D/linfor\_3D\_manual.pdf}.
Additionally, as we stated above, the 3D and 1D models used
in the computations shared identical atmospheric parameters, chemical
composition, equation of state, and opacities. We therefore tried to
minimize the differences in the model calculation procedure and line
synthesis computations, so that any discrepancy in the predictions
obtained with 3D and 1D models could be traced back to the differences
in physical realism invoked in the two types of models.

\begin{figure}[t]
\centering
\includegraphics[width=8.8cm]{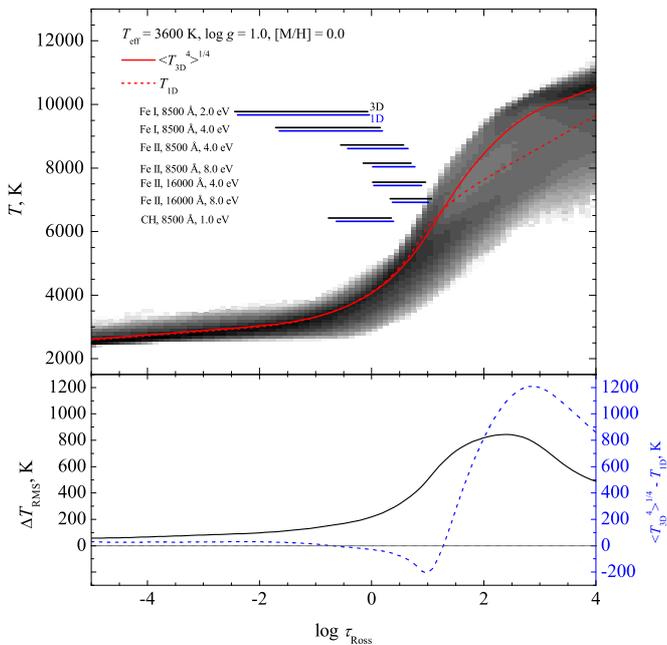}
\caption{Top panel: temperature profiles in the three model atmospheres
  of a red giant: 3D hydrodynamical (gray scales showing the logarithmic
  probability density of the 14-snapshot ensemble), average
  $\xtmean{\mbox{3D}}$ (red solid line, average over the 14-snapshot
  ensemble), and 1D LHD model atmosphere with \mlp=1.0 (red dashed line).
  Horizontal bars indicate approximate formation regions of \ion{Fe}{i},
  \ion{Fe}{ii}, and CH lines in the 3D and 1D models, at different wavelengths
  and line excitation potentials (bars mark the regions where 90\% of line
  equivalent width is acquired, i.e., between 5\% and 95\% in the cumulative
  line depression contribution function, see Appendix~\ref{sect:AB}).
  Bottom panel: RMS horizontal temperature fluctuations at constant
  \tauross in the 3D model (black solid line, 14-snapshot ensemble); and
  the difference between the temperature profiles corresponding to the
  average $\xtmean{\mbox{3D}}$ (14-snapshot ensemble average) and the
  1D model (blue dashed line).
\label{fig:T-struct-CFs}}
\end{figure}

\begin{figure}[t]
\centering
\includegraphics[width=8cm]{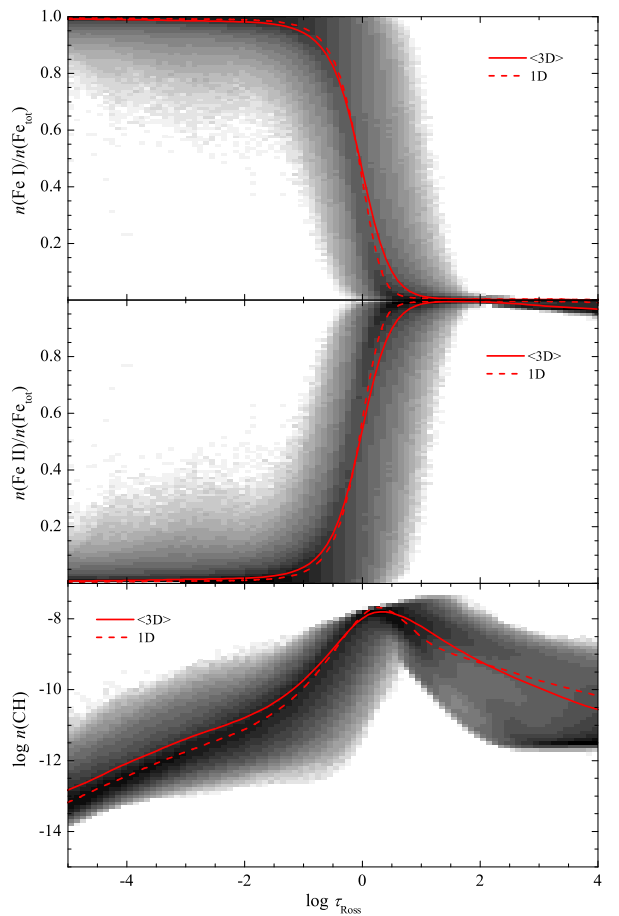}
\caption{Number densities of \ion{Fe}{i}, \ion{Fe}{ii}, and CH (top-down),
  plotted versus \tauross for the three model atmospheres of a red giant:
  3D hydrodynamical (gray scales showing the logarithmic probability density
  of the 14-snapshot ensemble), average $\xtmean{\mbox{3D}}$ (solid line,
  14-snapshot ensemble average), and 1D LHD model with \mlp=1.0 (dashed line).
  The number densities of \ion{Fe}{i} and \ion{Fe}{ii} are provided as
  fractions of the total iron number density, $n({\rm Fe_{tot}})$, whereas
  that of CH is given on a scale where $n({\rm H_{tot}})=12.0$.
\label{fig:numb-dens}}
\end{figure}

\subsection{3D--1D abundance corrections\label{sect:abund-corr-defin}}

The influence of convection on the spectral line formation and the resulting
line strengths was investigated with the aid of 3D--1D abundance corrections.
The 3D--1D abundance correction, $\Delta_{\rm 3D-1D}$, was defined as the
difference in the abundance $A({\rm X_{i}})$ of the element $\rm X_{i}$
obtained for a given equivalent width of a particular spectral line with the
3D hydrodynamical and classical 1D model atmospheres, $\Delta_{\rm
  3D-1D}=A({\rm X_{i}})_{\rm 3D} - A({\rm X_{i}})_{\rm 1D}$ \citep[see,
  e.g.,][]{CLS11}. The contribution to the 3D--1D abundance correction comes
from two major constituents: (a) the correction due to the horizontal
temperature fluctuations in the 3D model, $\Delta_{\rm
  3D-\langle3D\rangle}=A({\rm X_{i}})_{\rm 3D}-A({\rm X_{i}})_{\rm
  \langle3D\rangle}$, and (b) the correction due to differences between the
temperature profiles of the average $\xtmean{\mbox{3D}}$ and 1D models,
$\Delta_{\rm \langle3D\rangle-1D}=A({\rm X_{i}})_{\rm \langle3D\rangle}-A({\rm
  X_{i}})_{\rm 1D}$. The full abundance correction is a sum of the two
constituents, $\Delta_{\rm 3D-1D} =
\Delta_{\rm 3D-\langle3D\rangle}+\Delta_{\rm \langle3D\rangle-1D}$.

The 3D--1D abundance corrections were always calculated for weak lines
(equivalent width $<0.5$\,pm). The reason for this choice was that these
weak lines are supposed to be on the linear part of the curve-of-growth,
where their equivalent width is independent of the microturbulence velocity,
$\xi_{\rm mic}$, used with the $\xtmean{\mbox{3D}}$ and 1D models. Hence,
the derived 3D--1D abundance corrections become independent of the choice
of the microturbulence parameter.

\begin{figure*}[t]
\centering
\includegraphics[width=16cm]{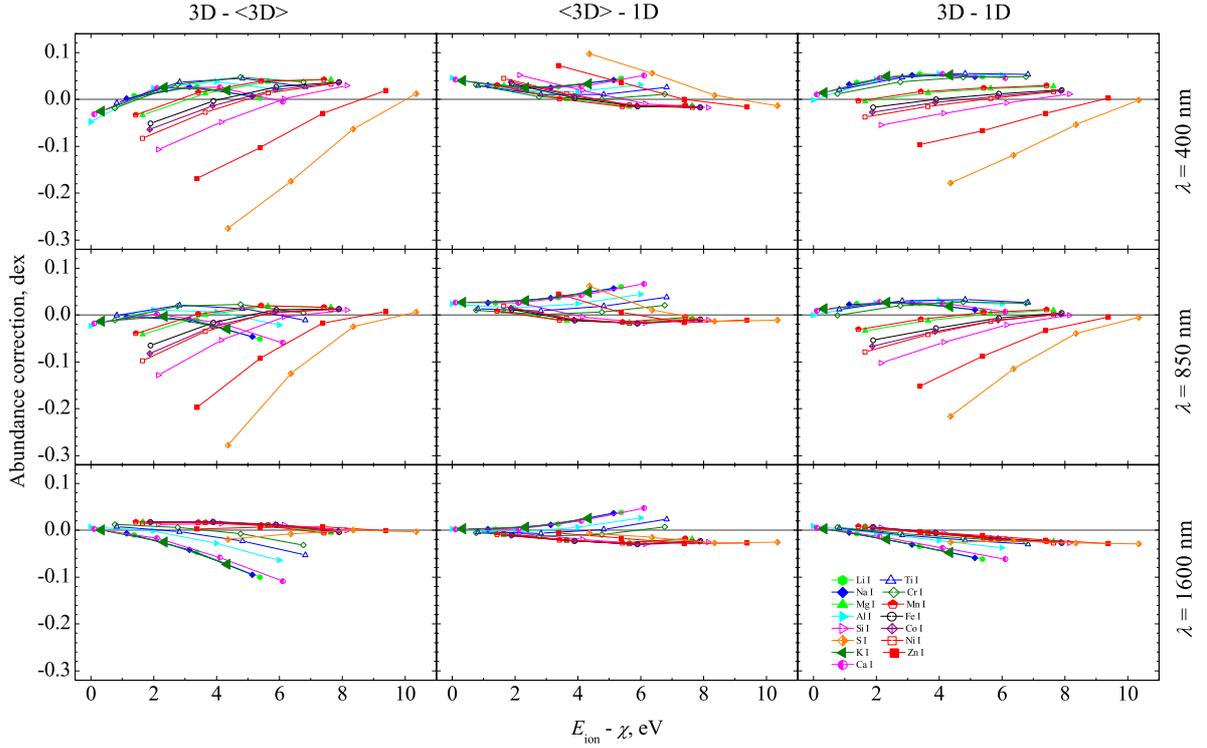}
\caption{Abundance corrections for spectral lines of neutral atoms, plotted
  versus the difference between their ionization energy and line excitation
  potential, $E_{\rm ion} - \chi$. Three types of abundance corrections are
  shown:  $\Delta_{\rm 3D-\langle3D\rangle}$ (left column), $\Delta_{\rm
    \langle3D\rangle-1D}$ (middle column), and $\Delta_{\rm 3D-1D}$ (right
  column). Abundance corrections are provided at three different wavelengths:
  400\,nm (top row), 850\,nm (middle row), and 1600\,nm (bottom row). The
  ionization energies of neutral atoms used in plotting
  this figure are provided in Table~\ref{table:e-ion}.
\label{fig:corr-atoms}}
\end{figure*}

\section{Results and discussion\label{sect:results}}

\subsection{General properties of spectral line formation in the red giant
atmosphere}

Convection indeed plays an important role in shaping the upper atmosphere
of the red giant studied here: convective up-flows and down-drafts dominate
close to the optical surface (optical depth $\tauross\approx1$), whereas
the shock-wave activity is most prominent in the outer atmosphere
\citep[see][]{LK12}. This alters the thermal structure of the
atmosphere, in particular, by causing spatial and temporal variations of
the temperature profiles in the 3D model (Fig.\,\ref{fig:T-struct-CFs}).
Since local temperature sets the physical conditions for spectral line
formation, it is reasonable to expect that line profiles predicted by
the 3D, $\xtmean{\mbox{3D}}$, and 1D models will also be different.

Obviously, there are significant differences between the number densities of
chemical species predicted by the 3D hydrodynamical and classical 1D models
(Fig.\,\ref{fig:numb-dens}). To a large extent, this behavior is defined by
the atomic/molecular properties of individual species, such as ionization
and dissociation potentials in case of atoms and molecules, respectively.
This leads to different sensitivities of the number densities to temperature
fluctuations. For example, species that are in the minority ionization stage
at a given depth in the atmosphere (e.g., \ion{Fe}{ii} at $\log\tauross\lesssim-0.2$)
are very sensitive to temperature fluctuations, since small changes in the
degree of ionization lead to a large spread in their number densities
(as indicated by the width of the density plot in Fig.\,\ref{fig:numb-dens}).

For a given set of spectral line parameters (such as wavelength, excitation
potential, and oscillator strength), the line formation region is
essentially defined by the temperature. Since differences between the
temperature profiles of the average $\xtmean{\mbox{3D}}$ and 1D models are
small in the entire range of optical depths relevant to the line formation
($|\Delta T|\lesssim200$\,K in the range of $\log \tau_{\rm Ross}\lesssim1.0$,
Fig.\,\ref{fig:T-struct-CFs}), this leads to very similar line formation
spans in the $\xtmean{\mbox{3D}}$ and 1D models. On the other hand, the
presence of horizontal temperature fluctuations in the line forming layers
plays a major role in the 3D line formation process, leading in general to
line strengthening with respect to the $\xtmean{\mbox{3D}}$ case. The
amplitude of the deviations from the average $\xtmean{\mbox{3D}}$
temperature profile, as defined by the RMS horizontal temperature fluctuations
($\Delta T_{\rm RMS} = \sqrt{\langle(T - T_0)^2\rangle_{x,y,t}}$, where
$\langle . \rangle_{x,y,t}$ denotes temporal and horizontal averaging on
surfaces of equal optical depth, and $T_0=\langle T \rangle_{x,y,t}$, is
the depth-dependent average temperature), is monotonically decreasing
throughout the entire photosphere, from $500$~K at $\log \tauross = +1.0$
to $50$~K at $\log \tauross=-5$ (Fig.\,\ref{fig:T-struct-CFs}). We thus
expect the differences between 3D and $\xtmean{\mbox{3D}}$ line formation
to show up most clearly for high-excitation lines, forming in the deep
photosphere where the horizontal temperature fluctuations are large. At
the same time, the differences between the average $\xtmean{\mbox{3D}}$
and 1D temperature profiles will also be most pronounced in this part
of the atmosphere, such that the $\xtmean{\mbox{3D}}$--1D effects should
also be strongest for the high-excitation lines.

\subsection{Abundance corrections for lines of neutral atoms}
\label{sect:acn}

\begin{table}[tb]
 \begin{center}
 \caption{Ionization energies of various neutral atoms.\label{table:e-ion}}
  \begin{tabular}{crcrcr}
  \noalign{\smallskip}
  \hline
   \noalign{\smallskip}
   Element & $E_{\rm ion}$, eV$^{\mathrm{a}}$ & Element & $E_{\rm ion}$, eV$^{\mathrm{a}}$ & Element & $E_{\rm ion}$, eV$^{\mathrm{a}}$ \\
    \hline\noalign{\smallskip}
    Li I  &  5.39 & S I   & 10.36 & Mn I  &  7.43 \\
    Na I  &  5.14 & K I   &  4.34 & Fe I  &  7.90 \\
    Mg I  &  7.65 & Ca I  &  6.11 & Co I  &  7.88 \\
    Al I  &  5.99 & Ti I  &  6.83 & Ni I  &  7.64 \\
    Si I  &  8.15 & Cr I  &  6.77 & Zn I  &  9.39 \\
  \hline
  \noalign{\smallskip}
  \end{tabular}
  \end{center}
\begin{list}{}{}
\item[$^{\mathrm{a}}$] NIST database, {\tt https://www.nist.gov}
\end{list}
\end{table}

\begin{figure*}[tb]
\centering
\includegraphics[width=16cm]{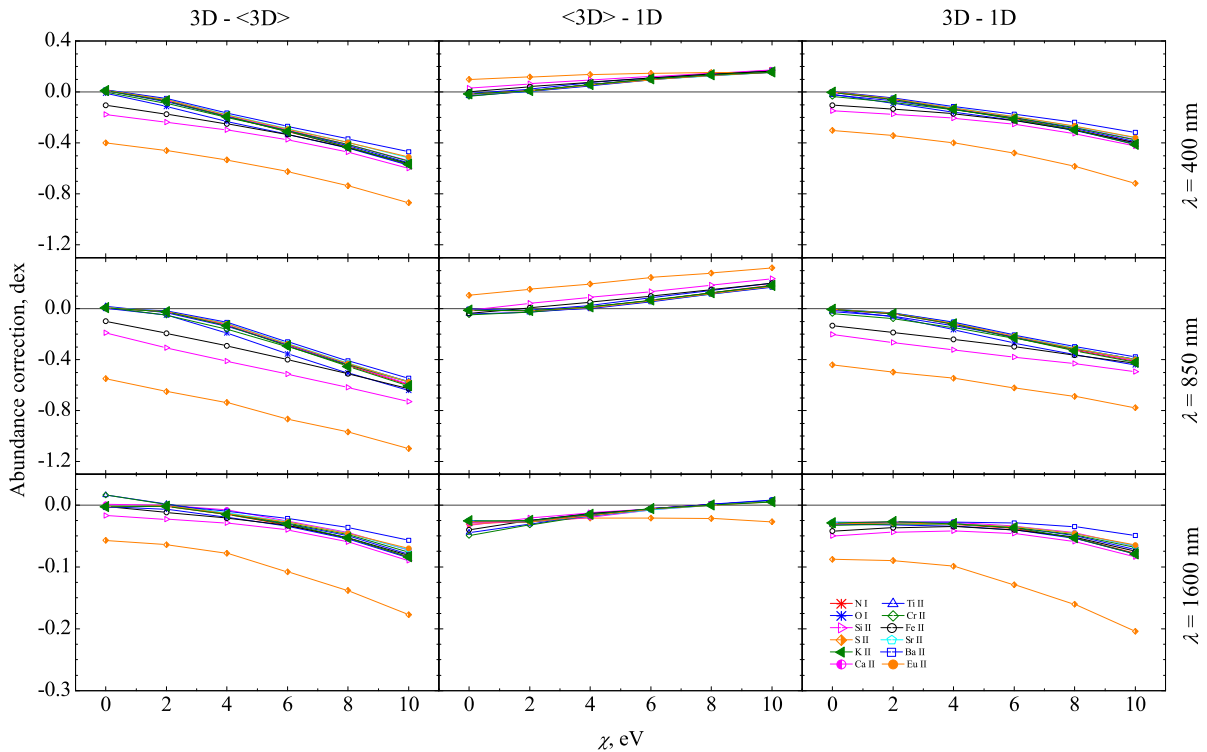}
\caption{Abundance corrections for spectral lines of ionized atoms (plus
  \ion{N}{i} and \ion{O}{i}), plotted versus the line excitation potential,
  $\chi$ (other notations as in Fig.\,\ref{fig:corr-atoms}).
\label{fig:corr-ions}}
\end{figure*}

The abundance corrections obtained for the bulk of neutral atoms are
plotted in Fig.\,\ref{fig:corr-atoms} versus the difference between their
ionization and excitation potentials, $E_{\rm ion} - \chi$. This choice of
abscissa was motivated by the fact that for chemical elements that
are almost completely ionized in the line-forming regions, it is the
\emph{difference} $E_{\rm ion} - \chi$ (and of course the thermodynamical
conditions) that defines the number density of the neutral atoms per unit
mass. The line opacity of this minority species is thus proportional to the
combined Saha-Boltzmann factor, $\kappa_\ell \sim \exp\{+(E_{\rm ion}-\chi)/kT\}$
\citep[cf.][in the analysis of the temperature dependence of the line 
strength]{Gray2005}. The abundance correction curves
shown in Fig.\,\ref{fig:corr-atoms} must therefore fall on top of each other
for all neutral atoms of elements that are strongly ionized (i.e.\ those with
sufficiently low ionization potential, $E_{\rm ion} \la 6$~eV), as explained
in more detail in Appendix\,\ref{sect:AA}. Indeed, this is clearly the case
for the abundance corrections of \ion{Li}{i}, \ion{Na}{i}, and \ion{K}{i}, which
are in their minority ionization stage throughout the entire atmosphere of
this particular red giant. For this type of atoms, the total abundance
correction, $\Delta_{\rm 3D-1D}$, and its constituents,
$\Delta_{\rm 3D-\langle3D\rangle}$ and $\Delta_{\rm \langle3D\rangle-1D}$,
are confined to the range of $-0.1\dots +0.05$\,dex, with comparable
contributions (of different sign) from the $\Delta_{\rm 3D-\langle3D\rangle}$
and $\Delta_{\rm \langle3D\rangle-1D}$ corrections. All corrections are more
negative in the near-IR at $\lambda\,1600$~nm than in the red at
$\lambda\,850$~nm.

For higher ionization potential, neutral atoms gradually turn
into majority species, and the abundance correction curves in
Fig.\,\ref{fig:corr-atoms} begin to separate. The two type of lines behave
radically different, and this can be clearly distinguished in the $\Delta_{\rm
  3D-1D}$ versus $E_{\rm ion} - \chi$ plot. This behavior is simply a consequence
of the combined action of ionization and excitation, as demonstrated in
Appendix\,\ref{sect:AA}. The point is that the ionization factor dominates
over the excitation factor as long as the neutral atoms are a
\emph{minority species}, and thus the line opacity decreases with increasing
temperature; low-excitation lines are then most temperature-sensitive. The
reverse is true for a neutral \emph{majority species}. Here the excitation
factor dominates over the ionization factor, and the line opacity increases
with increasing temperature, $\kappa_\ell \sim \exp\{- \chi/kT\}$
\citep[cf.][Chapter 13, Case 1]{Gray2005}. In this situation, high-excitation
lines are most temperature-sensitive.

The largest (most negative) 3D corrections are obtained for \ion{Zn}{i}
and \ion{S}{i}, reaching down to $\Delta_{\rm 3D-\langle3D\rangle}
\approx -0.20$ and $-0.28$~dex, respectively. This is explained by the fact
that these species have the highest ionization potentials of all atoms
shown in Fig.\,\ref{fig:corr-atoms}: the high-excitation lines of these atoms
have the most temperature-sensitive Boltzmann factor, and at the same time
form in the deep photosphere were the temperature fluctuations are more
pronounced than in the higher photospheric layers were the lines of the
minority species originate.  Surprisingly, the corrections for all
atoms of majority type are much smaller at $\lambda\,1600$~nm than at
$\lambda\,850$~nm (see Appendix~\ref{sect:AB} for a detailed explanation).
We note that neutral atoms with the highest ionization potentials
behave as the majority ions shown in Fig.\,\ref{fig:corr-ions}, where
\ion{N}{i} ($E_{\rm ion}=14.53$~eV) and \ion{O}{i} ($E_{\rm ion}=13.62$~eV)
have already been included. The abundance corrections for the ions (and
neutral majority species) show the same strong wavelength dependence.
They are discussed in the next section.

\begin{figure*}[tb]
\centering
\includegraphics[width=16cm]{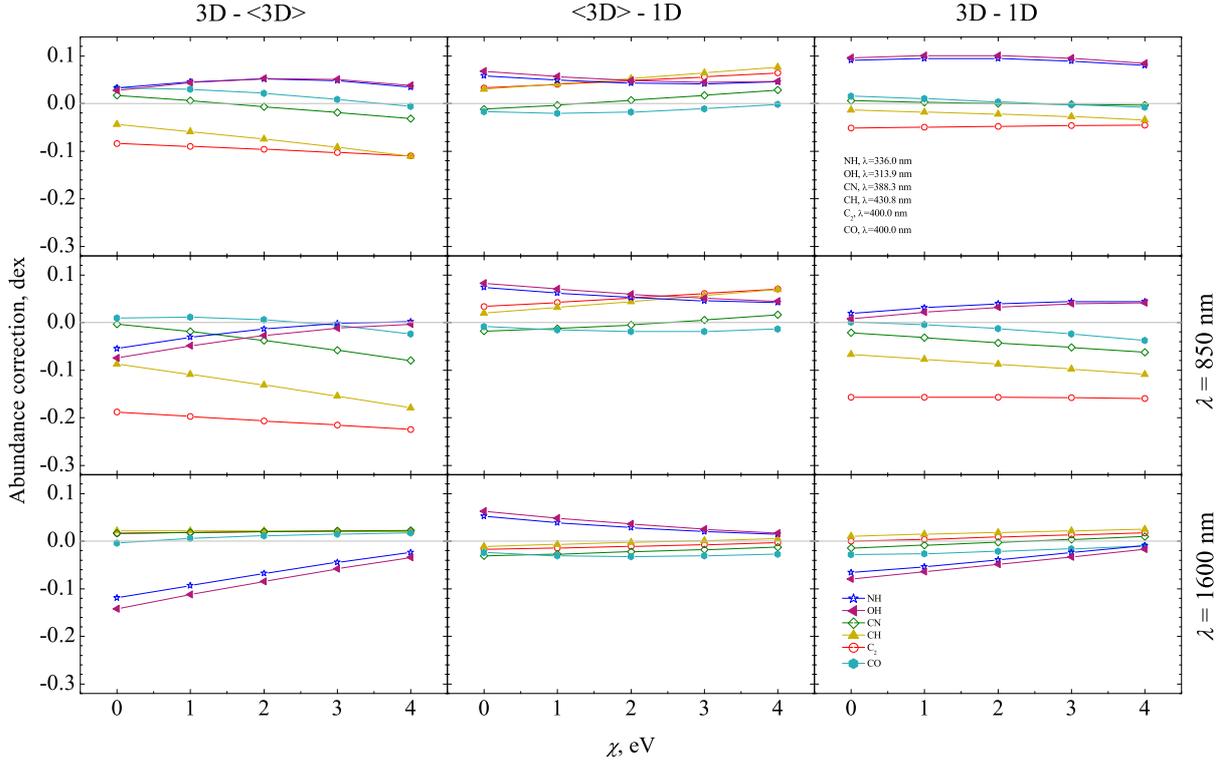}
\caption{Same as in Fig.\,\ref{fig:corr-ions} but for molecular lines. The
  bluest wavelength (top row) corresponds to the real wavelength of molecular
  bands in this spectral range (see legend), except for C$_2$ and CO, which do
  not have bands in the UV. The curve corresponding to C$_2$ coincides with
  that of CN in the lower left panel. Note the different $y$-scale in the
  lowest three panels.
\label{fig:corr-molec}}
\end{figure*}

\subsection{Abundance corrections for lines of ionized atoms}
\label{sect:aci}

Figure\,\ref{fig:corr-ions} displays the abundance corrections for a selection
of ionized atoms (plus \ion{N}{i} and \ion{O}{i}), showing a rather uniform
dependence on excitation potential $\chi$. We first consider the ions
that represent the majority ionization stage. Their line opacity (per unit
mass) depends on temperature as $\kappa_\ell \sim \exp\{- \chi/kT\}$. Owing
to its low first ionization potential, \ion{K}{ii} is present in its majority
ionization stage throughout the entire atmosphere of this red giant (\ion{K}{ii}
is only plotted for tracing the behavior of elements that
are nearly 100\% ionized; \ion{K}{ii} lines are inaccessible to
observations in real red giant atmospheres). The majority neutral atoms
\ion{N}{i} and \ion{O}{i} show exactly the same temperature sensitivity
(see Sect.\,\ref{sect:acn}), and hence their 3D--1D abundance corrections
coincide with those of the majority ions. The corrections vanish for ground-state
lines and increase steadily in amplitude toward the high-excitation
lines. At $\chi=10$~eV, $\Delta_{\rm 3D-\langle3D\rangle}$ amounts to
$\approx-0.6$~dex, while $\Delta_{\rm \langle3D\rangle-1D}$ reaches
$\approx+0.2$~dex, hence $\Delta_{\rm 3D-1D} \approx -0.4$~dex. For ions
that are no pure majority species, the abundance corrections depend weakly
on the ionization potential, systematically increasing in absolute size
with $E_{\rm ion}$.

Basically, the curves shown in Fig.\,\ref{fig:corr-ions} reflect the temperature
sensitivity of the line opacity, which simply scales with the excitation potential,
$\partial \log \kappa_\ell /\partial\log T = \chi/(k\,T)$ (valid for the majority
species). All abundance corrections are therefore close to zero for $\chi=0$,
and increase systematically with $\chi$. Since the $\xtmean{3D}$ model is cooler
in the line-forming regions than the 1D model, the same line is weaker in the
$\xtmean{3D}$ model than the 1D model, such that $\Delta_{\rm \langle3D\rangle-1D}
\sim -\chi\,(\log T_{\langle{\rm 3D}\rangle} - \log T_{\rm 1D})$ is positive.
On the other hand, $\Delta_{\rm 3D-\langle3D\rangle}$ is negative, because the
horizontally averaged line opacity in the 3D model exceeds the line opacity
obtained for the horizontally averaged temperature, $\langle\kappa_\ell\rangle >
\kappa_\ell(\langle T\rangle) \approx \kappa_\ell(T_{\langle\mathrm{3D}\rangle})$,
essentially because of the highly nonlinear temperature dependence of $\kappa_\ell$
($\partial^2\kappa_\ell/\partial T^2 >0$).

In this simple picture, the abundance corrections should be similar for all
wavelengths, since the temperature sensitivity of the line opacity is
wavelength independent (apart from a weak dependence through the stimulated
emission factor). Obviously, this is not the case: the abundance
corrections are much larger in the red at $\lambda\,850$~nm than in the
near-IR at $\lambda\,1600$~nm, even though the near-IR lines form at deeper
photospheric layers where the amplitude of the temperature fluctuations
should be  larger (Fig.\,\ref{fig:T-struct-CFs}, lower panel).
A detailed explanation of this counter-intuitive behavior is
given in Appendix \ref{sect:AB}. In short, it is not the line opacity alone
that determines the strength of a spectral line, but rather the ratio of
line-to-continuum opacity. It turns out that the continuum opacity at
$\lambda\,850$~nm (due to H$^-$ bound-free absorption) is almost
independent of temperature, such that the $T$-dependence of the line
opacity dominates the abundance corrections. At $\lambda\,1600$~nm,
however, the continuum opacity (due to H$^-$ free-free absorption) is strongly
$T$-dependent, leading to a substantial reduction of the temperature sensitivity
of the \emph{ratio} $\kappa_\ell/\kappa_{\rm c}$, and hence to much smaller
abundance corrections than at $\lambda\,850$~nm.

Finally, we note that the \ion{S}{ii} line exhibits by far the largest
abundance corrections, reaching $\Delta_{\rm 3D-\langle3D\rangle} \approx
-1.1$~dex at $\chi=10$~eV. This outstanding behavior is related to the
high ionization potential of \ion{S}{i} ($E_{\rm ion}=10.36$~eV),
making \ion{S}{ii} a true minority ion. In this case, the line opacity
is proportional to $\exp\{- (E_{\rm ion} + \chi)/kT\}$
\citep[cf.][Chapter 13, Case 3]{Gray2005}. This extremely high temperature
dependence gives rise to the extraordinary abundance corrections. In compliance
with this interpretation, $\Delta_{\rm 3D-\langle3D\rangle}$ for \ion{S}{ii}
at $\chi=0$ is already as large as $\Delta_{\rm 3D-\langle3D\rangle}$ at
$\chi=10$~eV for the normal majority ions. We note, however, that this is
probably a purely academic case, since such lines would certainly be too
weak to be observable.

\subsection{Abundance corrections for molecular lines}
\label{sect:acm}

Figure\,\ref{fig:corr-molec} displays the abundance corrections for a selection
of molecules as a function of the excitation potential $\chi$. The total abundance
corrections (right column) range from $\Delta_{\rm 3D-1D}\approx -0.16$\,dex
for C$_2$ at $\lambda\,850$~nm to $\approx +0.1$~dex for NH and OH in the
UV. As in the case of neutral atoms, the $\Delta_{\rm 3D-\langle3D\rangle}$
and $\Delta_{\rm \langle3D\rangle-1D}$ corrections are often of opposite sign,
making the total abundance correction, $\Delta_{\rm 3D-1D}$, slightly smaller
than the granulation correction $\Delta_{\rm 3D-\langle3D\rangle}$.

A detailed interpretation of the molecular abundance corrections is more
complicated than in the case of neutral atoms and ions, because the
dissociation equilibria of all molecules are closely coupled, such that
their number densities are intimately related and in some cases show an
unexpected temperature dependence. For basic orientation,
Fig.\,\ref{fig:frac-molec} (Appendix \ref{sect:AC}) shows the depth-dependence
of the molecule concentrations (number densities normalized to total number
density of carbon nuclei) in the 1D LHD model.

First of all, because of its high dissociation energy of
$D_0=11.1$~eV, CO is a majority species throughout the whole
photosphere; practically all carbon nuclei are locked in this molecule
at $\tauross < 1$, whereas CO dissociates quickly in the deeper photosphere.
In terms of abundance corrections, CO behaves like the  majority species
shown in  Fig.\,\ref{fig:corr-ions}. The dependence on excitation potential
is weaker, however. Owing to dissociation in the deeper layers, the excited CO
lines form higher up in the atmosphere than the majority ions with the same
$\chi$, and thus feel weaker temperature fluctuations and a reduced
$\xtmean{\rm 3D}$--1D temperature difference. As a consequence, the 3D
abundance corrections are small in the studied range of excitation potentials.

The molecules NH and OH belong to a different class of molecules. Their
constituents (H, N, and O) are majority species but only a tiny
fraction of N and O is bound in NH and OH molecules, respectively. In this
situation, the molecular line opacity is proportional to the Boltzmann
factor $\exp\{(D_0-\chi)/kT\}$, i.e.\ it depends on the difference
between dissociation energy $D_0$ and excitation potential $\chi$.
The largest abundance corrections are therefore found for $\chi=0$,
where the line opacity has the highest temperature sensitivity
(see Appendix \ref{sect:AC}, Eq.\,\ref{eqn:C5}). In fact, the abundance
corrections of these molecular lines behave very much like those of the
true minority neutral atoms shown in Fig,\,\ref{fig:corr-atoms}. This is
also true for their wavelength dependence.

The molecules CN, CH, and C$_2$ fall in yet another category. Their
temperature dependence is opposite to that of the other molecules,
i.e. their number density increases toward higher temperature (at
constant density or constant pressure). The reason is a strong coupling
to the dissociation equilibrium of CO, which controls the number density
of free carbon atoms. Higher temperatures lead to some dissociation of
CO, hence an increased concentration of carbon atoms, and in turn to
higher densities of CN, CH, and C$_2$. CH is more sensitive to this
coupling than CN, because the latter has a higher dissociation energy.
The C$_2$ molecule is most sensitive since its concentration depends
on the square of the number density of free carbon atoms. The abundance
corrections $\Delta_{\rm 3D-\langle3D\rangle}$ at $\lambda\,850$~nm are
therefore largest (most negative) for C$_2$.

At $\lambda\,1600$~nm, the fluctuation of the continuum opacity are
stronger than in the red at $\lambda\,850$~nm, because (i) the temperature
fluctuations are larger in the deeper layers where the near-IR lines form,
and (ii)  the temperature dependence\footnote{including the indirect
temperature dependence via electron pressure.} of the continuum opacity
is stronger in the near-IR (see Fig.\,\ref{fig:kappa_c}). For molecules like
NH and OH, the 3D--\xtmean{\mathrm{3D}} corrections are larger (more negative)
at $\lambda\,1600$~nm since $\kappa_\ell$ decreases while $\kappa_{\rm c}$
increases with $T$, such that the fluctuations of the ratio
$\kappa_\ell/\kappa_{\rm c}$ are stronger than the fluctuations of
$\kappa_\ell$. The molecules  CN, CH, and C$_2$, on the other hand,
have the opposite temperature dependence of $\kappa_\ell$, and thus
the fluctuations of the continuum opacity effectively cancel the
fluctuations of the line opacity. The corrections
$\Delta_{\rm 3D-\langle3D\rangle}$ for  CN, CH, and C$_2$ are therefore
much smaller in the near-IR than in the red.

\subsection{Abundance corrections at $2.5\,\mu$m}
\label{sect:a25}

As a test case, we also calculated the 3D--1D corrections for several
fictitious lines of different neutral, ionized atoms and molecules at
$2.5\,\mu$m (not shown in Figs.~\ref{fig:corr-atoms}-\ref{fig:corr-molec}). In
this case the $\Delta_{\rm 3D-\langle3D\rangle}$, $\Delta_{\rm
  \langle3D\rangle-1D}$, and $\Delta_{\rm 3D-1D}$ abundance corrections, as
well as their dependence on the line parameters, were very similar to those
obtained at $1.6\,\mu$m. This result is hardly surprising since opacity in the
red giant atmospheres at $1.6-2.6\,\mu$m is dominated by ${\rm H}^{-}$
free-free transitions, the efficiency of which varies very little in this
wavelength range. Thus, the line formation regions are essentially the same at
both wavelengths, which leads to very similar line formation properties and
3D--1D abundance corrections.

\subsection{Line formation in the 1D model atmospheres: influence of the
mixing-length parameter $\mlp$\label{sect:alphamlt}}

According to the classical Schwarzschild criterion, 1D \LHD\ models of the
red giant studied here are convectively stable in the outer photospheric
layers ($\log \tau_{\rm Ross}\lesssim1$), irrespective of the mixing length
parameter, $\mlp$ \citep[][]{KHL05,LK12}. Since line formation depths
typically never reach deeper than $\log \tau_{\rm Ross}\approx0.6$, the
choice of \mlp\ should have no effect on the resulting spectral line
strengths. The only exceptions may occur in case of near-infrared lines
and/or lines characterized by very high excitation potentials ($>8$\,eV),
i.e.\ those that form deepest in the photosphere and thus may be sensitive
to differences in the temperature profiles computed with different
mixing-length parameters.

To verify whether this can actually happen, we ran several test calculations
using fictitious \ion{Fe}{i} (0, 4\,eV), \ion{Fe}{ii} (0, 4, 8\,eV), and
\ion{O}{ii} (0, 10\,eV) lines located at 850\,nm and 1600\,nm. We found
that differences between the abundances derived using model atmospheres
calculated with different mixing-length parameters ($\mlp=1,2$) were always
very small. In fact, they never exceeded $0.02$\,dex and thus may be safely
ignored in most situations related with the abundance work in this
particular red giant.

\subsection{How to use the 3D abundance corrections}

We recall that in this work the 3D--1D corrections were calculated
for very weak spectral lines. While such corrections provide a convenient way
to estimate the importance of 3D hydrodynamical effects in the spectral
line formation, abundance corrections for stronger lines may be different
because of saturation effects, leading to differences in formation depths and
introducing a sensitivity to velocity fields (see Appendix\,\ref{sect:AB_sat}).

The theoretical 3D abundance corrections were derived from the
comparison of the equivalent width of the same (artificial) spectral line
computed with a 3D model and a 1D model. Since the 3D and
the 1D models use the same stellar parameters, atomic data, and numerical
methods (as far as possible), the resulting \emph{differential} 3D
corrections should be applicable to any 1D model, irrespective
of the physical details.

For given stellar parameters (\Teff, \logg, \moh) and a given element and
ionization stage, the 3D abundance correction depends only on the energy
of the lower level of the transition, the wavelength of the line, and in
principle also on the mixing-length parameter used for the 1D model,
$\Delta_{\mathrm{3D}-\mathrm{1D}}(\chi, \lambda, \mlp)$, provided that
the line is \emph{weak} (on the linear part of the curve-of-growth). In
this case, the corrections given in Figs.\,\ref{fig:corr-atoms},
\ref{fig:corr-ions}, and \ref{fig:corr-molec} can be readily applied to
the 1D LTE abundance determinations performed with any standard 1D
mixing-length model atmosphere: $A({\rm X_{i}})_{\rm 3D} =
A({\rm X_{i}})_{\rm 1D} + \Delta_{\mathrm{3D}-\mathrm{1D}}$. As shown in
Sect.\,\ref{sect:alphamlt}, the dependence on \mlp\ is negligible for the
present stellar parameters.

In the general case of stronger lines, the 3D corrections depend
in addition on the the equivalent width of the line, $W$, and on the microturbulence
parameter used for the 1D model, $\xi_{\rm mic}$,
$\Delta_{\mathrm{3D}-\mathrm{1D}}(\chi, \lambda, W, \xi_{\rm mic}; \mlp)$,
as illustrated in Fig.\,\ref{fig:ac_ew} for \ion{Fe}{ii}, $\chi=10$~eV.
Except for this example, we do not provide the dependence of
$\Delta_{\mathrm{3D}-\mathrm{1D}}$ on $W$ and $\xi_{\rm mic}$ in this
work. The given corrections for the weak line limit can then only serve as an
indication of the line's susceptibility to 3D effects. Figure \ref{fig:ac_ew}
suggests that the 3D correction for stronger lines is always more positive
than in the weak line limit.  The figure also seems to indicate that
$\Delta_{\mathrm{3D}-\mathrm{1D}}$ becomes independent of $W$ for strong
lines, suggesting that a `strong line limit' might be a useful quantity.
Additional investigations are needed to see whether these properties remain
valid in general.

\section{Conclusions\label{sect:conclusions}}

We have investigated the influence of thermal convection on the spectral line
formation in the atmosphere of a red giant located close to the RGB tip
($\Teff=3660$, $\log g=1.0$, $\moh=0.0$). To this end, we synthesized a large
number of fictitious atomic and molecular lines of astrophysically important
tracer elements, using for this purpose the 3D hydrodynamical \COBOLD\ and
classical 1D \LHD\ stellar model atmospheres (both types of models sharing
identical atmospheric parameters, opacities, equation of state, and chemical
composition). The influence of convection on the line formation was
investigated by focusing on the differences between the abundances inferred
from a given (weak) spectral line by the requirement of producing a given
 equivalent width with the 3D and 1D model atmospheres, assuming LTE.

Overall, convection plays a considerable role in the atmosphere of the red
giant studied here. There are significant horizontal temperature fluctuations
seen at different optical depths in the atmosphere that are caused either by
convective up- or down-flows (inner atmosphere), convective overshoot,
and/or shock wave activity (outer atmosphere). This leads to substantial
horizontal variations in the number densities of certain chemical species,
causing significant deviations from the predictions of a classical 1D model.

Spectral line formation is more or less strongly affected by
convection in this particular red giant, depending on the line parameters
and the chemical species under consideration. The differences in elemental
abundances inferred from a given spectral line with the 3D and 1D model
atmospheres, $\Delta_{\rm 3D-1D}$, are most pronounced for high-excitation
lines of ions and atoms of predominantly neutral elements. Their abundance
corrections grow larger with increasing excitation potential, reaching values
of $\Delta_{\rm 3D-1D}\sim -0.4$~dex for excitation potential $\chi=10$~eV.
The main physical reason for this 3D--1D difference is the increasingly
nonlinear temperature dependence of the line opacity as $\chi$ increases.
In a 3D atmosphere, the horizontal temperature fluctuations then lead to an
enhancement of the effective line opacity with respect to the 1D case.

Lines of neutral atoms of predominantly ionized elements show significantly
smaller corrections, with $\Delta_{\rm 3D-1D}$ not exceeding $\pm0.1$\,dex.
Here the temperature dependence of the line opacity due to ionization and
excitation tend to cancel partially.

Molecular lines show a more complex behavior due to the strong coupling
between the different dissociation equilibria. CO is least susceptible to
3D effects, with $|\Delta_{\rm 3D-1D}|<0.03$~dex. NH and OH represent a
different category of molecules; their abundance corrections fall in the
range $|\Delta_{\rm 3D-1D}|<0.1$~dex. The largest corrections are found for
C$_2$, reaching as low as $\Delta_{\rm 3D-1D}\approx -0.16$~dex.

It is important to emphasize that the 3D--1D corrections for all chemical species
studied here show a significant wavelength dependence. In most cases,
the abundance corrections are significantly smaller in the near-infrared at
$\lambda\,1600$~nm than in the optical spectral range.
Careful investigation reveals that this does \emph{not} indicate that the
atmospheric layers where the infrared lines originate are least affected
by convection. Rather, the strong wavelength dependence of the 3D corrections
is related to the fact that the continuum opacity is much more temperature
dependent around $\lambda\,1600$~nm (mainly H$^-$ free-free absorption)
than around  $\lambda\,850$~nm (mainly H$^-$ bound-free absorption).
It should be stressed though that this conclusion was reached for
the particular red giant studied here and may change with the atmospheric
parameters of the star under investigation.

One may thus conclude that the spectral line formation in a red giant
photosphere is a delicate process, governed by the subtle interplay
between microscopic (atomic line parameters) and macroscopic (local
temperature fluctuations) physics. The size of the 3D-hydrodynamical
fluctuations and their effects on the line formation process can only be
assessed by dedicated radiation hydrodynamics simulations and detailed
3D line formation calculations. For the present example, we find
substantial 3D abundance corrections, especially for ionized atoms,
suggesting that even at solar metallicity, the accuracy of
1D LTE stellar abundance work can be significantly improved by exploiting
the additional information provided by realistic 3D model atmospheres.

\begin{acknowledgements}

We thank the referee R. Collet for a comprehensive and constructive
report, which helped to improve the paper. This work was supported by grant
from the Research Council of Lithuania (MIP-101/2011). HGL acknowledges
financial support from EU contract MEXT-CT-2004-014265 (CIFIST), and by the
Sonderforschungsbereich SFB\,881 ``The Milky Way System'' (subproject A4)
of the German Research Foundation (DFG). AK and HGL acknowledge financial
support from the the Sonderforschungsbereich SFB\,881 ``The Milky Way System''
(subproject A4) of the German Research Foundation (DFG) that allowed exchange
visits between Vilnius and Heidelberg. PB and AK acknowledge support from the
Scientific Council of the Observatoire de Paris and the Research Council of
Lithuania (MOR-48/2011) that allowed exchange visits between Paris and Vilnius.
MS acknowledges funding from the Research Council of Lithuania for a
research visit to Vilnius.
\end{acknowledgements}

\bibliographystyle{aa}

\begin{thebibliography}{}

\bibitem[{Bessell} {et al.}(2004)]{BCG04}
Bessell,~M., Christlieb,~N., Gustafsson,~B.
2004, \apjl, 612, 61

\bibitem[{Caffau} {et al.}(2008)]{CLS08}
Caffau,~E., Ludwig,~H.-G., Steffen,~M., Ayres,~T.R., Bonifacio,~P., Cayrel,~R., Freytag,~B., \& Plez,~B.
2008, \aap, 488, 1031

\bibitem[{Caffau} {et al.}(2011)]{CLS11}
Caffau,~E., Ludwig,~H.-G., Steffen,~M., Freytag,~B., and Bonifacio,~P.
2011, SoPh, 268, 255

\bibitem[{Collet} {et al.}(2007)]{CAT07}
Collet, R., Asplund, M., \& Trampedach, R.
2007, A\&A, 469, 687

\bibitem[Cox(2000)]{AQ2000} Cox, A.N.\ 2000, Allen's
Astrophysical Quantities, 4th ed.

\bibitem[{Dobrovolskas} {et al.}(2010)]{DKL10}
Dobrovolskas,~V., Ku\v{c}inskas,~A., Ludwig,~H.-G., Caffau,~E., Klevas,~J., \& Prakapavi\v{c}ius,~D.
2010, Proc. of 11th Symposium on Nuclei in the Cosmos, Proceedings of Science, ID~288 ({\tt arXiv:1010.2507})

\bibitem[{Freytag} {et al.}(2012)]{FSL12}
Freytag,~B., Steffen,~M., Ludwig,~H.-G., Wedemeyer-Bhm,~S., Schaffenberger,~W., \& Steiner,~O.
2012, J. Comp. Phys., 231, 919

\bibitem[{{Gratton}(1985)}]{G85}
Gratton,~R.
1985, \aap, 148, 105

\bibitem[Gray(2005)]{Gray2005} Gray, D.~F.\ 2005,
The Observation and Analysis of Stellar Photospheres,
3rd Edition, Cambridge University Press

\bibitem[{Grevesse} \& {Sauval}(1998)]{GS98}
Grevesse,~N., \& Sauval,~A.J.
1998, \ssr, 85, 161

\bibitem[{Gustafsson} {et al.}(2008)]{GEK08}
Gustafsson,~B., Edvardsson,~B., Eriksson,~K., J{\o}rgensen,~U.G., Nordlund,~{\AA}., \& Plez,~B.
2008, \aap, 486, 951

\bibitem[{Ivanauskas} {et al.}(2010)]{IKL10}
Ivanauskas,~A., Ku\v{c}inskas,~A., Ludwig,~H.-G., \& Caffau,~E.
2010, Proc. of 11th Symposium on Nuclei in the Cosmos, Proceedings of Science, ID~290 ({\tt arXiv:1010.1722})

\bibitem[{Ku\v{c}inskas} {et al.}(2005)]{KHL05}
Ku\v{c}inskas,~A., Hauschildt,~P.H., Ludwig,~H.-G., Brott,~I., Vansevi\v{c}ius,~V., Lindegren,~L., Tanab\'{e},~T., \& Allard,~F.
2005, \aap, 442, 281

\bibitem[{Lambert}(1968)]{L68}
Lambert,~D.
1968, \mnras, 138, 143

\bibitem[{Ludwig}(1992)]{L92}
Ludwig,~H.-G.
1992, Ph.D. Thesis, Univ. Kiel

\bibitem[{Ludwig \& Ku\v{c}inskas}(2012)]{LK12}
Ludwig, H.-G. \& Ku\v{c}inskas, A.
2012, \aap, in press

\bibitem[{Ludwig} {et al.}(1994)]{LJS94}
Ludwig,~H.-G., Jordan,~S., \& Steffen,~M.
1994, \aap, 284, 105

\bibitem[{Ludwig} {et al.}(2009)]{LCS09}
Ludwig,~H.-G., Caffau,~E., Steffen,~M., Freytag,~B., Bonifacio,~P., \& Ku\v{c}inskas,~A.
2009, \memsai, 80, 711

\bibitem[{Magain}(1986)]{M86}
Magain,~P.
1986, \aap, 163, 135

\bibitem[{Mihalas}(1978)]{M78}
Mihalas,~D.
1978, Stellar Atmospheres, Freeman and Company

\bibitem[{Nordlund}(1982)]{N82}
Nordlund,~{\AA}.
1982, \aap, 107, 1

\bibitem[{Spite} {et al.}(2005)]{SCP05}
Spite,~M., Cayrel,~R., Plez,~B., Hill,~V., Spite,~F., Depagne,~E., Fran\c{c}ois,~B., Bonifacio,~P., Barbuy,~B., Beers,~T.,
Andersen,~J., Molaro,~P., Nordstr{\o}m,~B., and Primas,~F.
2005, \aap, 430, 655

\bibitem[{Steffen} \& {Holweger}(2002)]{SH02}
Steffen,~M. and Holweger,~H.
2002, \aap, 387, 258

\bibitem[{Steffen} {et al.}(1995)]{SLF95}
Steffen,~M., Ludwig,~H.-G., \& Freytag,~B.
1995, \aap, 300, 473

\bibitem[{V\"{o}gler}(2004)]{VBS04}
V\"{o}gler,~A., Bruls,~J.H.M.J., \& Sch\"{u}ssler,~M.
2004, \aap, 421, 741


\end{thebibliography}

\begin{appendix}

\section{Analysis of the abundance corrections for lines of neutral atoms}
\label{sect:AA}

The temperature sensitivity of spectral lines originating from the neutral
atoms of partially ionized species (e.g.\ Mg\,{\sc i}, Ca\,{\sc i},
Fe\,{\sc i}) is governed by the Saha and Boltzmann equations, i.e.\ by
changes of the degree of ionization and of the excitation of the line's
lower level. In the following, we develop a simplified model of the
3D--$\langle$3D$\rangle$ abundance corrections, which result from horizontal
fluctuations of the thermodynamical quantities.

\begin{figure}[tb]
\centering
\mbox{\includegraphics[bb=32 48 570 380, width=8.9cm]{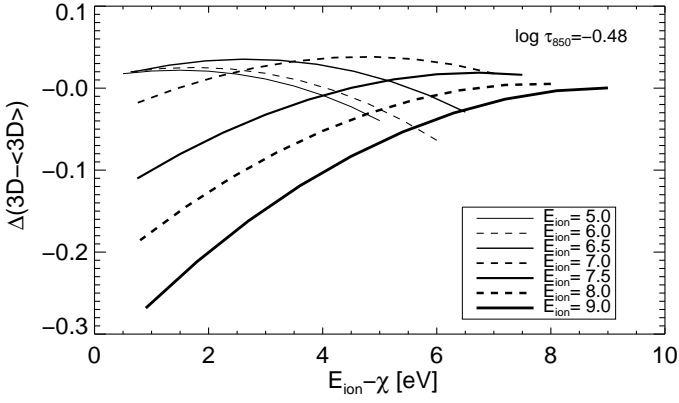}}
\caption{Abundance correction $\Delta_{\rm 3D-\langle{3D}\rangle}$,
  computed according to Eqs.\,(\ref{eqn:AA1}) -- (\ref{eqn:AA4}),
  versus the difference between ionization and excitation potential,
  $E_{\rm ion}-\chi$. Each curve corresponds to a different value of
  $E_{\rm ion}$. The thermodynamic variables $\theta(x,y)$ and $p_e(x,y)$
  were taken from the 3D model at monochromatic optical depth
  $\log \tau_{850} = -0.48$, where the mean temperature is $3360 \pm 140$~K.
  The temperature dependence of the partition functions $U_0$ and $U_1$
  has been neglected.}
\label{fig:kappa_l}
\end{figure}

The ratio of the total number of neutral atoms $n_0$ (with ionization energy
$E_{\rm ion}$ and partition function
$U_0$) to the total number of singly ionized atoms $n_1$
(with partition function $U_1$) at temperature $T=5040/\theta$ and electron
pressure $p_e$ is given by Saha's equation
\citep[see e.g.][Eq.\,1.20]{Gray2005}
\begin{eqnarray}
\log \frac{n_0}{n_1} &=&  E_{\rm ion}\,\theta +2.5\,\log \theta
                         + \log {p_e} + \log \frac{U_0(\theta)}{U_1(\theta)}
                         -9.08 \nonumber \\
&\equiv& \log f(\theta,p_e).
\label{eqn:AA1}
\end{eqnarray}
The opacity of a line transition with lower level excitation potential $\chi$
is proportional to the number of absorbing atoms (per unit mass) in this
state of excitation,
\beq
\kappa_\ell \sim  \frac{n_0}{n_0+n_1}\,10^{-\theta\,\chi} =
\frac{1}{1+f^{-1}(\theta,p_e)}\,10^{-\theta\,\chi}\, .
\label{eqn:AA2}
\eeq
In the presence of horizontal fluctuations of $\theta$ and $p_e$, the average
line opacity is amplified with respect to the $\xtmean{\mbox{3D}}$
case by a factor
\beq
\mathcal{A}_\ell(\tauc) = \frac{\langle\kappa_\ell(\theta,p_e)\rangle_{x,y}}
{\kappa_\ell\left(\langle\theta\rangle_{x,y},
\langle{p_e}\rangle_{x,y}\right)}\, ,
\label{eqn:AA3}
\eeq
where $\langle.\rangle_{x,y}$ denotes horizontal averaging at constant
continuum optical depth $\tauc$. Assuming that fluctuations of the continuum
opacity and the source function can be neglected, the abundance correction
for weak lines can be estimated as
\beq
\Delta_{\rm 3D-\langle{3D}\rangle} \approx -\log \mathcal{A}_\ell\, .
\label{eqn:AA4}
\eeq
Figure \ref{fig:kappa_l} shows the result of computing
$\Delta_{\rm 3D-\langle{3D}\rangle}$ according to
Eqs.\,(\ref{eqn:AA1}) -- (\ref{eqn:AA4}) for different
combinations of $\chi$ and $E_{\rm ion}$. The figure illustrates
how the curves $\Delta_{\rm 3D-\langle{3D}\rangle}$ versus $E_{\rm ion}-\chi$
change systematically as the parameter $E_{\rm ion}$ increases from
$0$ to $9$~eV. For $E_{\rm ion} \la 6$~eV, the neutral atoms are a minority
species, $f \ll 1$, and we see from Eq.\,(\ref{eqn:AA2}) that in this case
\beq
\kappa_\ell \sim f\,10^{-\theta\,\chi} \sim 10^{\theta\,(E_{\rm ion}-\chi)}
\label{eqn:AA5}
\eeq
depends only on the difference $(E_{\rm ion}-\chi)$, such that all curves
with  $E_{\rm ion} \la 6$~eV fall on top of each other. As $E_{\rm ion}$
increases further, the vertex of the curves moves from
$(E_{\rm ion}-\chi)_{\rm max} =0$~eV to $\approx 4$~eV to $9$~eV as the
ionization balance shifts from $\langle f\rangle_{x,y} \approx 0$
($E_{\rm ion}=0$~eV) to $\approx 1$ ($E_{\rm ion}=7$~eV) to $\ga 1000$
($E_{\rm ion}=9$~eV). In fact, it can be shown analytically that
\beq
(E_{\rm ion}-\chi)_{\rm max} \approx E_{\rm ion}\,\frac{\langle f\rangle_{x,y}}
{1 + \langle f\rangle_{x,y}}\, .
\label{eqn:AA6}
\eeq

\begin{figure}[tb]
\centering
\mbox{\includegraphics[bb=32 32 570 380, width=8.9cm]{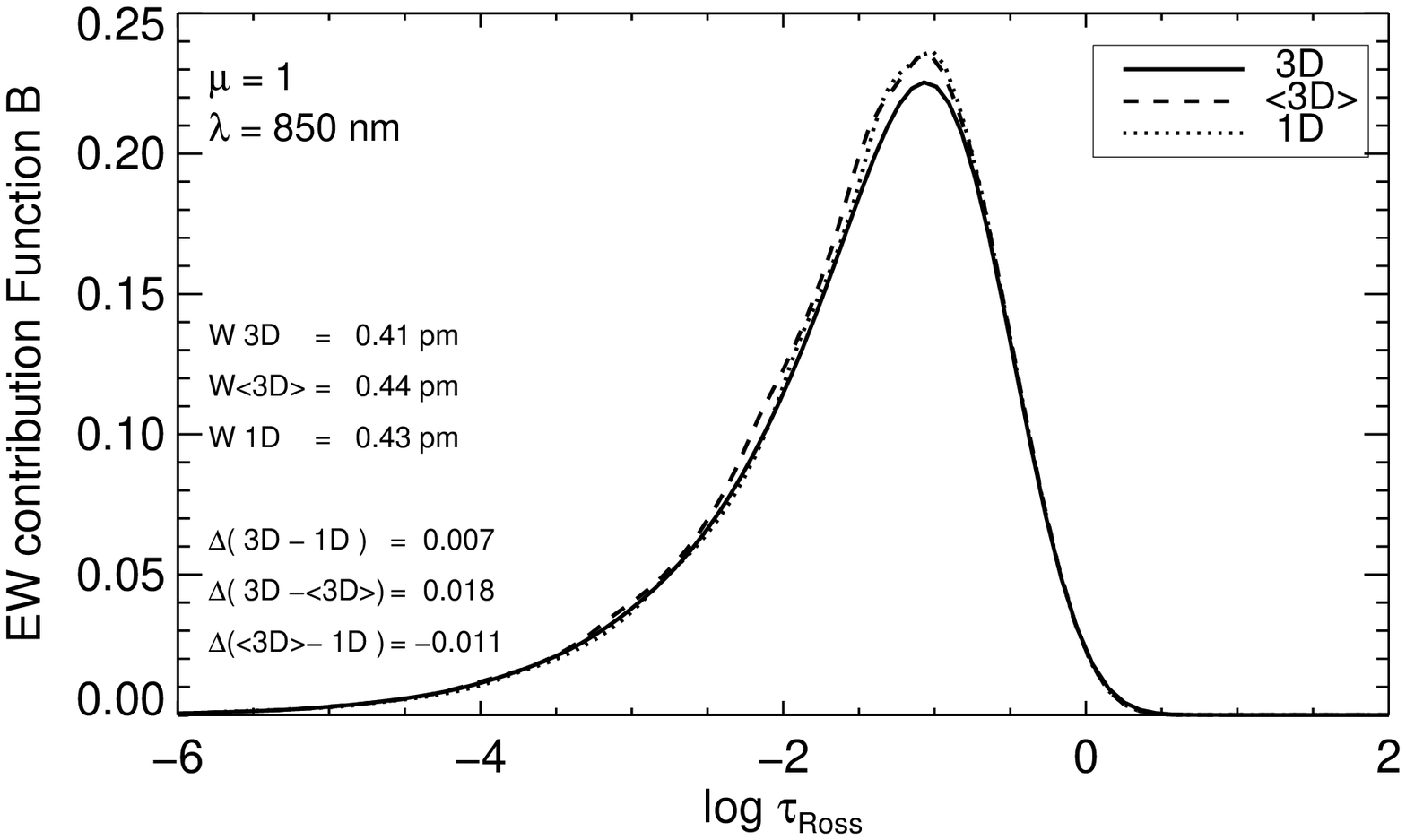}}
\mbox{\includegraphics[bb=32 32 570 380, width=8.9cm]{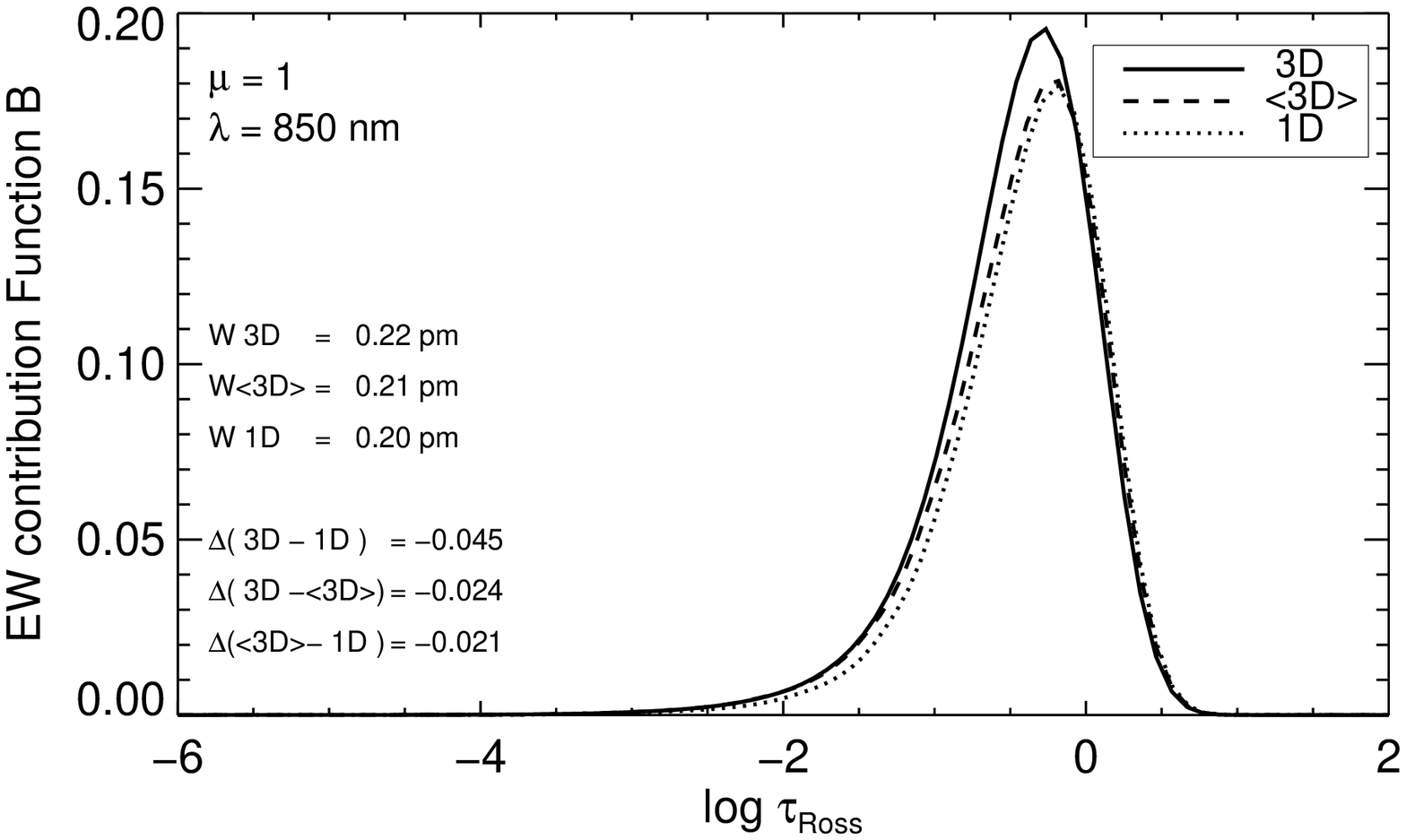}}
\caption{Disk-center ($\mu=1$) equivalent width contribution functions,
  $\mathcal{B}(\log \tauc)$, of a weak (artificial) \ion{Fe}{i} line with
  excitation potential $\chi=0$~eV (top) and $\chi=5$~eV (bottom),
  at wavelengths $\lambda\,850$~nm, evaluated according to
  the weak line approximation (Eqs.\,\ref{eqn:B5}, \ref{eqn:B7}),
  for a single snapshot of the 3D model, the corresponding
  $\langle$3D$\rangle$ average model, and the associated 1D LHD model
  used in this work. The contribution functions, originally defined on
  the monochromatic optical depth scale, have been transformed to the
  Rosseland optical depth scale.}

\label{fig:cfFeI}
\end{figure}

Admittedly, the description of the abundance corrections developed
above is severely simplified. It ignores the fact that the line formation
region is extended and that the location of its center of gravity
depends sensitively on  $E_{\rm ion}$ and $\chi$ (cf.\ Figs.\,\ref{fig:cfFeI},
\ref{fig:cfFeII}, \ref{fig:cfmol}). Also, fluctuations of
the continuum opacity and the source function were neglected. Nevertheless,
the systematics seen in Fig.\,\ref{fig:kappa_l} provides a basic explanation
of the detailed numerical results presented in Sect.\,\ref{sect:acn},
Fig.\,\ref{fig:corr-atoms} (especially middle left panel).

\section{Analysis of the abundance corrections for high-excitation
lines of ions}
\label{sect:AB}

In the following, we analyze in some detail the abundance corrections
derived for the high-excitation \ion{Fe}{ii} lines, which are
representative of the ionized atoms and show the largest 3D corrections
(see Fig.\,\ref{fig:corr-ions}). Evaluating the
\emph{equivalent width contribution functions} of this line in the 3D
and the 1D models, we determine the physical cause
of the abundance corrections. In particular, we can understand the sign
of the 3D--$\langle$3D$\rangle$ and $\langle$3D$\rangle$--1D corrections
and explain why these corrections are so much smaller at
$\lambda\,1600$~nm than at $\lambda\,850$~nm.

For simplicity, we consider only vertical rays (disk-center intensity)
in a single snapshot from the 3D simulation, noting that the qualitative
behavior of the abundance corrections is similar for intensity and flux,
and does not vary much in time, i.e. the 3D--$\langle$3D$\rangle$ correction
is always strongly negative, while the $\langle$3D$\rangle$--1D correction is
slightly positive at $\lambda\,850$~nm.

\subsection{Formalism}
\label{sect:AB_form}

Following \citet{M86}, the \emph{line depression
contribution function} (for vertical rays, LTE), $C_I^D$, is defined
in \LINFOR\ as
\begin{eqnarray}
C_I^D(\Delta\lambda,\,\tauc) =
\left\langle\,\eta(\Delta\lambda,\tauc)\,\uclam(\tauc) \,
\exp\{-\left(\tauc + \taul(\Delta\lambda,\tauc)\right)\}\,
\right\rangle_{x,y} \nonumber \\
\mathrm{}\, ,
\label{eqn:B1}
\end{eqnarray}
where $\tauc$ is the continuum optical depth, $\eta(\Delta\lambda,\tauc) =
\kappa_\ell(\Delta\lambda,\tauc)/\kappa_{\rm c}(\tauc)$ is the ratio of line
opacity to continuum opacity,
$\uclam(\tauc) = I_{\rm c}^+(\tauc) - S_{\rm c}(\tauc)$ is the difference
between outgoing continuum intensity and source
function\footnote{$\uclam$ is also called `source function gradient'
since $\uclam=\mathrm{d}S_{\rm c}/\mathrm{d}\tauc$ in the diffusion
approximation, and if $S_{\rm c}$ is a linear function of $\tauc$.}, and
$(\tauc + \taul(\Delta\lambda))$ is the total optical depth in the line;
angle brackets $\left\langle{.}\right\rangle_{x,y}$ indicate horizontal
averaging at constant continuum optical depth. Note that $\eta$ (and $\taul$)
vary with wavelength position in the line profile, $\Delta\lambda$, whereas
$\tauc$ and $\uclam$ can be considered as constant across the line profile.
Then the absolute line depression at any wavelength in the line profile is
\beq
D_I(\tauc=0,\Delta\lambda) =
\int_0^\infty\, C_I^D(\Delta\lambda,\,\tauc')\, \diff{\tauc'}\, .
\label{eqn:B2}
\eeq
Defining further the \emph{equivalent width contribution function} as
\beq
C_I^W(\tauc) = \int_{-\infty}^{+\infty} C_I^D(\Delta\lambda',\,\tauc)\,
 \diff{\Delta\lambda'}\, ,
\label{eqn:B3}
\eeq
the equivalent width of the line is finally computed as
\begin{eqnarray}
W_I &=& \frac{1}{\langle I_{\rm c}\rangle}\,
\int_0^\infty\, C_I^W(\tauc')\, \diff{\tauc'} \nonumber \\
&=& \frac{1}{\langle I_{\rm c}\rangle}\,
\int_{-\infty}^\infty\, \mathcal{B}(\log \tauc')\, \diff{\log \tauc'}\, ,
\label{eqn:B4}
\end{eqnarray}
where $\langle I_{\rm c}\rangle$ is the horizontally averaged emergent
continuum intensity, and $\mathcal{B}$ is defined as
\beq
\mathcal{B}(\tauc) = \ln 10\,\tauc\,C_I^W(\tauc)\, .
\label{eqn:B5}
\eeq

In the limit of \emph{weak lines}, we can assume that $\taul \ll \tauc$ over
the whole line formation region, and Eqs.\,(\ref{eqn:B1}) and (\ref{eqn:B3})
simplify to
\beq
C_I^D(\Delta\lambda,\,\tauc) = \exp\{-\tauc\}\,
\left\langle\,\eta(\Delta\lambda,\tauc)\,\uclam(\tauc)\,\right\rangle_{x,y}\, ,
\label{eqn:B6}
\eeq
and
\beq
C_I^W(\tauc) = \exp\{-\tauc\}\,
\left\langle\,\eta_0(\tauc)\,\uclam(\tauc)\,\right\rangle_{x,y}\, ,
\label{eqn:B7}
\eeq
where
\beq
\eta_0(\tauc) = \int_{-\infty}^{+\infty} \eta(\Delta\lambda',\,\tauc)\,
 \diff{\Delta\lambda'}\, .
\label{eqn:B8}
\eeq
In this weak line limit, $C_I^D$, and hence $C_I^W$ are strictly proportional
to the line opacity, and the equivalent width scales linearly with the
$gf$-value of the line (or the respective chemical abundance). Then the
3D abundance corrections can simply be obtained from the equivalent widths
as
\begin{eqnarray}
\Delta_{\rm 3D-1D} \;\; &=& -\log\,(W_{\rm 3D}/W_{\rm 1D})\, , \nonumber \\
\Delta_{\rm 3D-\langle\mathrm{3D}\rangle} &=&
-\log\,(W_{\rm 3D}/W_{\rm \langle\mathrm{3D}\rangle})\, ,
\nonumber \\
\Delta_{\rm \langle\mathrm{3D}\rangle - \mathrm{1D}} &=&
-\log\,(W_{\langle\mathrm{3D}\rangle}/W_{\rm 1D})\, .
\label{eqn:B9}
\end{eqnarray}

\subsection{Analysis of mixed contribution functions}
\label{sect:AB_form_cfm}

\begin{figure}[tb]
\centering
\mbox{\includegraphics[bb=32 32 570 380, width=8.9cm]{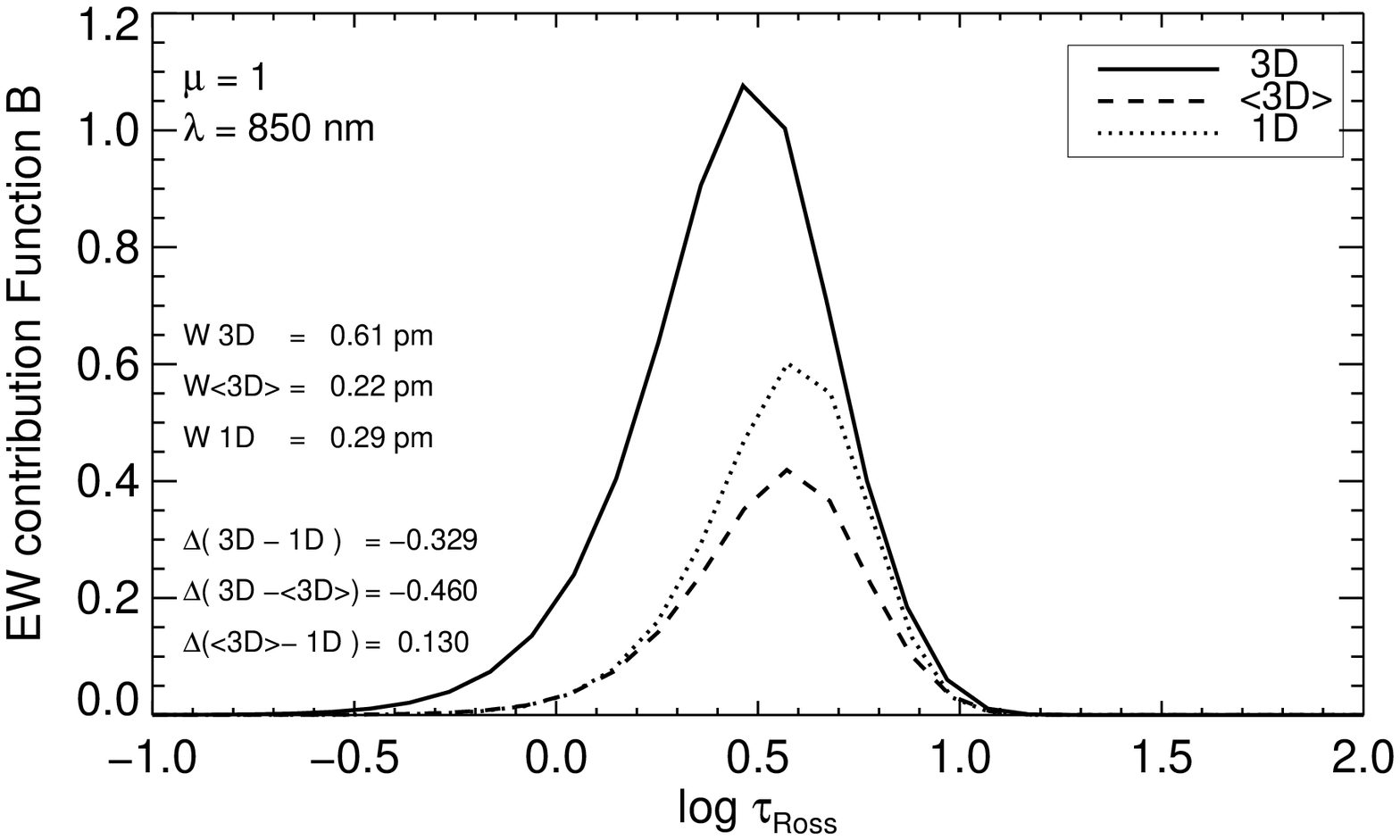}}
\mbox{\includegraphics[bb=32 32 570 380, width=8.9cm]{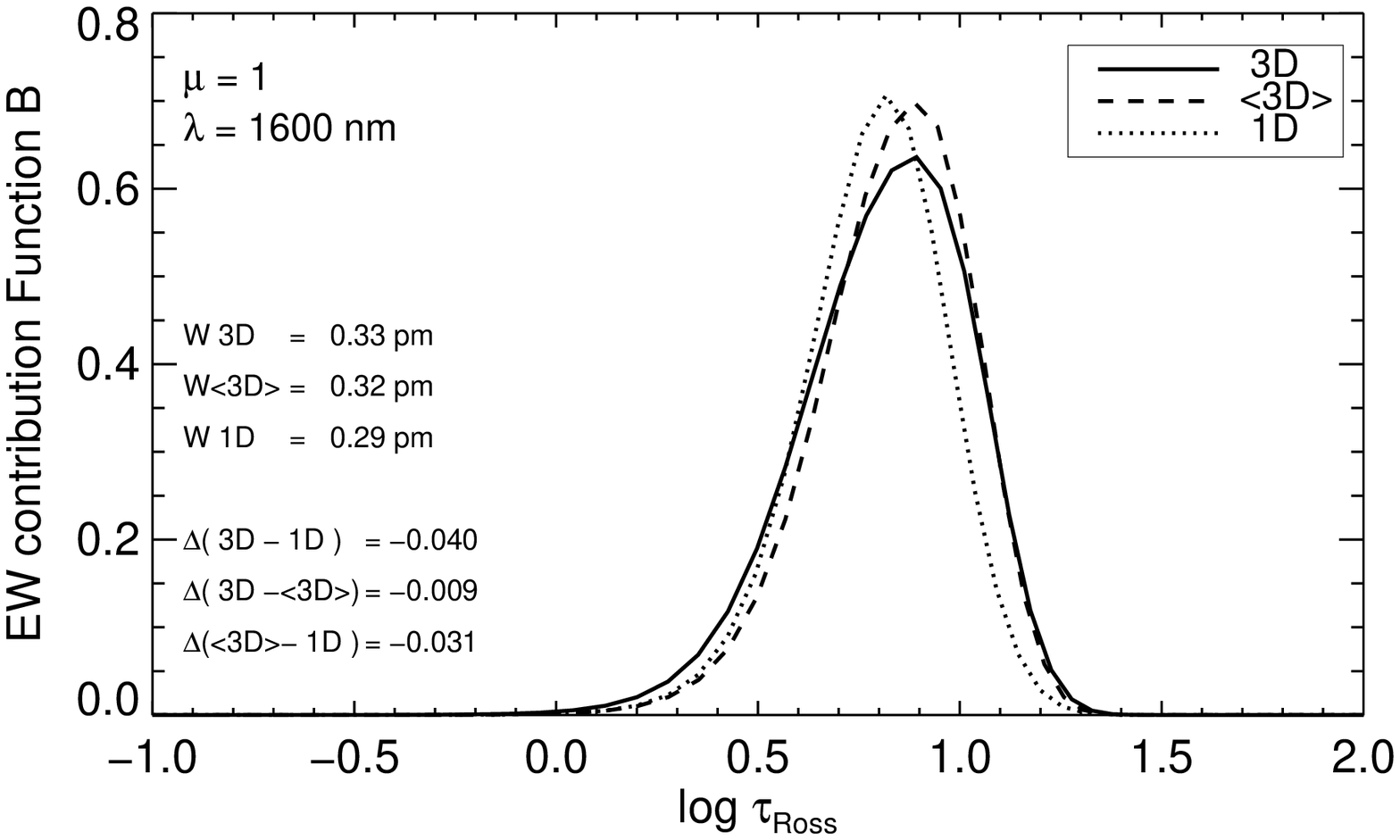}}
\caption{Disk-center ($\mu=1$) equivalent width contribution functions,
  $\mathcal{B}(\log \tauc)$, of a weak (artificial) \ion{Fe}{ii} line with
  excitation potential $\chi=10$~eV, at wavelengths $\lambda\,850$~nm (top) and
  $\lambda\,1600$~nm (bottom), evaluated according to
  the weak line approximation (Eqs.\,\ref{eqn:B5}, \ref{eqn:B7}),
  for a single snapshot of the 3D model, the corresponding
  $\langle$3D$\rangle$ average model, and the associated 1D LHD model used in
  this work. The contribution functions, originally defined on the
  monochromatic optical depth scale, have been transformed to the Rosseland
  optical depth scale as a common reference.}
\label{fig:cfFeII}
\end{figure}

Figure \ref{fig:cfFeII} compares the equivalent width contribution functions,
$\mathcal{B}$, of a weak (artificial) \ion{Fe}{ii} line with
excitation potential $\chi=10$~eV at two wavelengths, $\lambda\,850$~nm
(top) and $\lambda\,1600$~nm (bottom), for the 3D model, the
corresponding $\langle$3D$\rangle$ average model, and the associated 1D LHD
model used in this work. For each of the different models the
area below the corresponding curve is proportional to the equivalent width
of the emerging line profile. At $\lambda\,850$~nm, the equivalent width
produced by the 3D model is significantly larger than that of the 1D LHD
model, which in turn is significantly lager than that of the
$\langle$3D$\rangle$ model, assuming the same iron abundance in all cases.
The abundance corrections derived with Eq.\,(\ref{eqn:B9}) are
$\Delta_{\rm 3D-1D}=-0.33$, $\Delta_{\rm 3D-\langle 3D\rangle}=-0.46$, and
$\Delta_{\rm \langle 3D\rangle -1D}=+0.13$~dex. These numbers are fully
consistent with the results shown in Fig.\,\ref{fig:corr-ions}.
At $\lambda\,1600$~nm, on the other hand, the equivalent widths
obtained from the three different model atmospheres are obviously very similar;
all abundance corrections are much smaller than those at $\lambda\,850$~nm,
again in basic agreement with the results shown in Fig.\,\ref{fig:corr-ions}.

Figure \ref{fig:cfFeII} also shows that the formation region of this
high-excitation \ion{Fe}{ii} line is well confined to a narrow region in the
deep photosphere, mainly below $\tauross=1$, where the degree of ionization
of iron is changing rapidly with depth. As expected, the line originates from
somewhat deeper layers at $\lambda\,1600$~nm (minimum of H$^-$ opacity)
than at $\lambda\,850$~nm (maximum of H$^-$ opacity). According to
Fig.\,\ref{fig:T-struct-CFs}, both the amplitude of the horizontal temperature
fluctuations and the 1D - $\langle$3D$\rangle$ temperature difference
increase with depth in the range $-1 < \log \tauross < +1$, such that they
are larger in the line formation region at $\lambda\,1600$~nm. Naively,
one would thus expect the amplitude of the abundance corrections
$\Delta_{\rm 3D-\langle 3D\rangle}$ and $\Delta_{\rm \langle 3D\rangle -1D}$
to be \emph{larger} at $\lambda\,1600$~nm than at $\lambda\,850$~nm.
However, this reasoning obviously fails. As we have seen before, the abundance
corrections are found to be strikingly \emph{smaller} at $\lambda\,1600$~nm.
Additional analysis is necessary to resolve this apparent contradiction.

To understand the origin of the abundance corrections, we need to understand
the role of the different factors that make up the contribution function
$\mathcal{B}$, essentially $u_{\rm c}$ and $\eta_0 = \kappa_\ell / \kappa_{\rm c}$.
The different behavior of these factors in the different types of models
determines the sign and amplitude of the abundance corrections.
In the following, subscripts $1$, $2$, and $3$ refer to the 1D LHD model,
the $\langle$3D$\rangle$ model, and the 3D model, respectively. With
this notation in mind, we define the \emph{mixed contribution functions}
\beq
\mathcal{B}_{i,j,k}\,(\tauc) = \ln 10\,\tauc\,\exp\{-\tauc\}\,
\left\langle\,\uclam{_{,i}}\,(\tauc)\,
\frac{\kappa_{\ell, j}\,(\tauc)}{\kappa_{{\rm c}, k}\,(\tauc)\,}
\right\rangle_{x,y}\, ,
\label{eqn:B10}
\eeq
where $i= 1 \ldots 3$, $j= 1 \ldots 3$, $k= 1 \ldots 3$. The three mixed
contribution functions with three identical subscripts $i=j=k$ are thus the
\emph{normal} contribution functions for the 1D, $\langle$3D$\rangle$,
and 3D model, respectively. From each of the $\mathcal{B}_{i, j, k}$ we can
compute an equivalent width according to Eq.\,(\ref{eqn:B4}), which we
denote as $W_{i, j, k}$. The equivalent widths can then be used to derive
abundance corrections via
\beq
\Delta_{i,j,k,m} = -\log (W_{i,j,k}/W_{m,m,m})\, .
\label{eqn:B11}
\eeq
The numerical evaluation of the relevant abundance corrections is compiled in
Table \ref{table:dijkm}.

\begin{table}
 \begin{center}
 \caption{Abundance corrections for the \ion{Fe}{ii} line ($\chi=10$~eV)
 derived from mixed contribution functions $\mathcal{B}_{i,j,k}$ at
 $\lambda\,850$ and $1600$~nm.}
  \begin{tabular}{ccrrcrr}
  \noalign{\smallskip}
  \hline
   \noalign{\smallskip}
   Case & $i$\,$j$\,$k$\,$m$ & $\Delta_{i,j,k,m}$ & $\Delta_{i,j,k,m}$ &
          $i$\,$j$\,$k$\,$m$ & $\Delta_{i,j,k,m}$ & $\Delta_{i,j,k,m}$ \\
        &                 & 850~nm        & 1600~nm &
                          & 850~nm        & 1600~nm \\
    \hline\noalign{\smallskip}
0 & 2\,2\,2\,2 &  0.000 &  0.000 & 1\,1\,1\,1 &  0.000 &  0.000 \\
1 & 3\,2\,2\,2 &  0.014 &  0.036 & 2\,1\,1\,1 &  0.062 & -0.053 \\
2 & 2\,3\,2\,2 & -0.451 & -0.146 & 1\,2\,1\,1 &  0.140 &  0.150 \\
3 & 3\,3\,2\,2 & -0.558 & -0.165 & 2\,2\,1\,1 &  0.202 &  0.093 \\
4 & 2\,2\,3\,2 &  0.003 & -0.132 & 1\,1\,2\,1 & -0.077 & -0.130 \\
5 & 3\,2\,3\,2 &  0.046 & -0.040 & 2\,1\,2\,1 & -0.015 & -0.184 \\
6 & 2\,3\,3\,2 & -0.374 & -0.015 & 1\,2\,2\,1 &  0.069 &  0.026 \\
7 & 3\,3\,3\,2 & -0.460 & -0.009 & 2\,2\,2\,1 &  0.130 & -0.031 \\
\hline
  \noalign{\smallskip}
  \end{tabular}
  \end{center}
\label{table:dijkm}
\end{table}

\subsubsection{$\langle$3D$\rangle$--1D abundance corrections}
\label{sect:AB_ac1}

With the help of Table \ref{table:dijkm}, the physical interpretation of
the $\Delta_{\langle\mathrm{3D}\rangle-\mathrm{1D}}$ abundance correction is
straightforward. Columns (6) and (7) show the effect of the
different factors that contribute to the $\langle$3D$\rangle$--1D abundance
correction. Owing to the different thermal structure of the two model
atmospheres, all three factors, $u_{\rm c}$, $\kappa_\ell$, and $\kappa_{\rm  c}$
change simultaneously, and the full correction is
$\Delta_{\langle\mathrm{3D}\rangle-\mathrm{1D}}=\Delta_{2,2,2,1}$, listed
in the last row of the Table as case (7). The other cases (1)--(6) refer to
`experiments' where only one or two of the factors are allowed to change
while the remaining factors are fixed to expose the abundance
corrections due to the individual factors. Case (1), for example, shows the
correction that would result for fixed opacities, $\kappa_\ell(\tauc)$,
$\kappa_{\rm c}(\tauc)$, accounting only for the differences in the source
function gradient $u_{\rm c}$. Case (6) shows the complementary experiment
where $u_{\rm c}$ is fixed and both opacities are changing in accordance with
the different thermodynamical conditions.

At $\lambda$\,$850$~nm, the continuum opacity is dominated by H$^-$
bound-free absorption, which shows its maximum at this wavelength. The
high-excitation \ion{Fe}{ii} line forms around $\log\tau_{850}\approx +0.7$
($\log\tauross\approx +0.6$), i.e.\ significantly below continuum
optical depth unity. At this depth,  both the temperature and
the temperature gradient are slightly lower in the
$\langle\mathrm{3D}\rangle$ model than in the 1D model. As a consequence,
both $u_{\rm c}$ and $\kappa_{\rm c}$ decrease toward the
$\langle\mathrm{3D}\rangle$ model, approximately by the same factor (see cases
1 and 4), and hence their effects cancel out. The highly temperature-dependent
line opacity ($\partial{\log \kappa_\ell}/\partial{\theta} \approx -10$;
$\theta=5040/T$) is thus the dominating factor and determines the total
$\Delta_{\langle\mathrm{3D}\rangle-\mathrm{1D}}$ abundance correction
(compare cases 2 and 7).

\begin{figure}[tb]
\centering
\mbox{\includegraphics[bb=32 48 570 380, width=8.9cm]{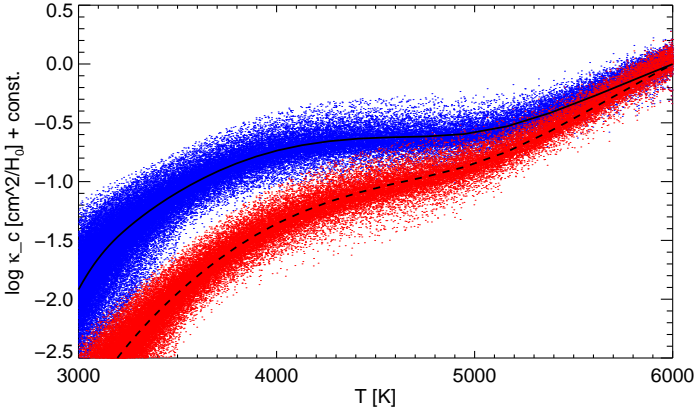}}
\caption{Continuous opacity due to H$^-_{\rm bf}$ at $\lambda\,850$~nm
(solid line, cloud of blue dots), and due to H$^-_{\rm ff}$ at
$\lambda\,1600$~nm (dashed line, cloud of red dots),
as a function of temperature in the
3D model (dots) and in the $\langle$3D$\rangle$ model (lines), respectively.
The opacities have been computed according to Eqs.\,(8.12) and (8.13) by
\cite{Gray2005}; they are given in units of cm$^2$ per neutral hydrogen atom,
and have been normalized to $1$ at $T=6000$~K. Both opacities are
proportional to the electron pressure, and their ratio is a simple
monotonic function of temperature. Note that $\kappa($H$^-_{\rm bf})$
is almost temperature insensitive between $T=4000$ and $5000$~K.}
\label{fig:kappa_c}
\end{figure}

At $\lambda$\,$1600$~nm, the situation is different. Here
the continuum opacity is mainly due to H$^-$ free-free absorption.
The important difference is that the temperature sensitivity of the
H$^-$ free-free opacity is significantly higher than that of the H$^-$
bound-free absorption (see Fig.\,\ref{fig:kappa_c}).
In the line formation region around $\log\tau_{1600}\approx +0.15$
($\log\tauross\approx +0.85$), the temperature sensitivity of
$\kappa_\ell$ and $\kappa_{\rm c}$ is now comparable, such that the ratio of
both opacities is nearly the same in the two models. The corrections due to
$\kappa_\ell$ and $\kappa_{\rm c}$ are almost equal and of opposite sign
(cases 2 and 4), and hence cancel out. At the same time, the source function
gradient is very similar in both models, and thus the correction due to
$u_{\rm c}$ is small (case 1). The total $\Delta_{\langle\mathrm{3D}\rangle-\mathrm{1D}}$
abundance correction is therefore significantly smaller than at
$\lambda$\,$850$~nm (case 7).

\subsubsection{3D--$\langle$3D$\rangle$ abundance corrections}
\label{sect:AB_ac2}

The physical interpretation of the
$\Delta_{\mathrm{3D}-\langle\mathrm{3D}\rangle}$ abundance correction proceeds
along similar lines. Columns (3) and (4) of Table\,\ref{table:dijkm}
show the influence of the different factors that contribute to the
3D--$\langle$3D$\rangle$ `granulation correction'. The three factors,
$u_{\rm c}$, $\kappa_\ell$, and $\kappa_{\rm c}$ differ between the 3D and the
$\langle$3D$\rangle$ model due to the presence of horizontal fluctuations
of the thermodynamical conditions at constant optical depth $\tauc$,
which then lead to more or less \emph{nonlinear} fluctuations of the
factors that make up the contribution function. The full correction,
$\Delta_{\mathrm{3D}-\langle\mathrm{3D}\rangle}=\Delta_{3,3,3,2}$,
allows for fluctuations in all three factors and is listed in the last
row of the table as case (7). The other cases (1)--(6) refer to `experiments'
where the fluctuations are artificially suppressed for one or two of the
factors to study the impact of the fluctuations of the individual factors
on the resulting the abundance correction. For example, case (5) shows the
correction that would result if the fluctuations of the line opacity,
$\kappa_\ell(\tauc)$, were suppressed. case (2) shows the complementary
experiment where only $\kappa_\ell(\tauc)$ is allowed to fluctuate, while
$u_{\rm c}$ and $\kappa_{\rm c}$ are fixed.

At $\lambda$\,$850$~nm, our weak \ion{Fe}{ii} line
is strongly enhanced in the 3D model due to the nonlinear fluctuations
of the line opacity. The fluctuations lead to a line enhancement, and
hence to \emph{negative} 3D--$\langle$3D$\rangle$ abundance corrections,
whenever $\langle\kappa_\ell(T)\rangle_{x,y} >
\kappa_\ell(\langle T\rangle_{x,y})$, which happens to be the case as
$\kappa_\ell \propto \exp\{-E/kT\}$ (roughly speaking because
$\partial^2 \kappa_\ell/\partial T^2>0$).
As can be deduced from the comparison of cases (2) and (7), suppression of the
fluctuations of both $u_{\rm c}$ and $\kappa_{\rm c}$ does not change the
resulting 3D abundance correction.  We can furthermore see that the fluctuations
of $u_{\rm c}$ enhance the nonlinearity of the fluctuations of $\kappa_\ell$
(case 3) and that the fluctuations of $\kappa_{\rm c}$ diminish the
nonlinearity of the fluctuations of $\kappa_\ell$ (case 6). We conclude that
the fluctuations of $u_{\rm c}$ and $\kappa_{\rm c}$ must be substantial, but
essentially linear, such that they do not produce any significant abundance
corrections on their own (cases 1, 4, and 5).

At $\lambda$\,$1600$~nm, the continuum opacity $\kappa_{\rm c}$ is lower
than at $\lambda$\,$850$~nm, and our weak \ion{Fe}{ii} line forms at
somewhat deeper layers where the temperature is higher. Equally important,
the temperature sensitivity of $\kappa_{\rm c}$ is distinctly higher at
$\lambda$\,$1600$~nm than at $\lambda$\,$850$~nm, as is demonstrated
in Fig.\,\ref{fig:kappa_c}.
This fact is the key to understanding the drastically smaller abundance
corrections found at $\lambda$\,$1600$~nm.

Comparing the effect of the line opacity fluctuations for the two wavelengths
(case 2), we see that the corresponding abundance correction is significantly
smaller at $\lambda$\,$1600$~nm. This result is unexpected, because
according to Fig.\,\ref{fig:T-struct-CFs} the temperature fluctuations,
$\delta T_{\rm rms}$, ought to be larger in the deeper layers where the near-IR
line forms, which in turn should lead to more nonlinear fluctuations of
the line opacity and hence larger abundance corrections at
 $\lambda$\,$1600$~nm compared to $\lambda$\,$850$~nm.

\begin{figure}[tb]
\centering
\mbox{\includegraphics[bb=32 32 570 380, width=8.9cm]{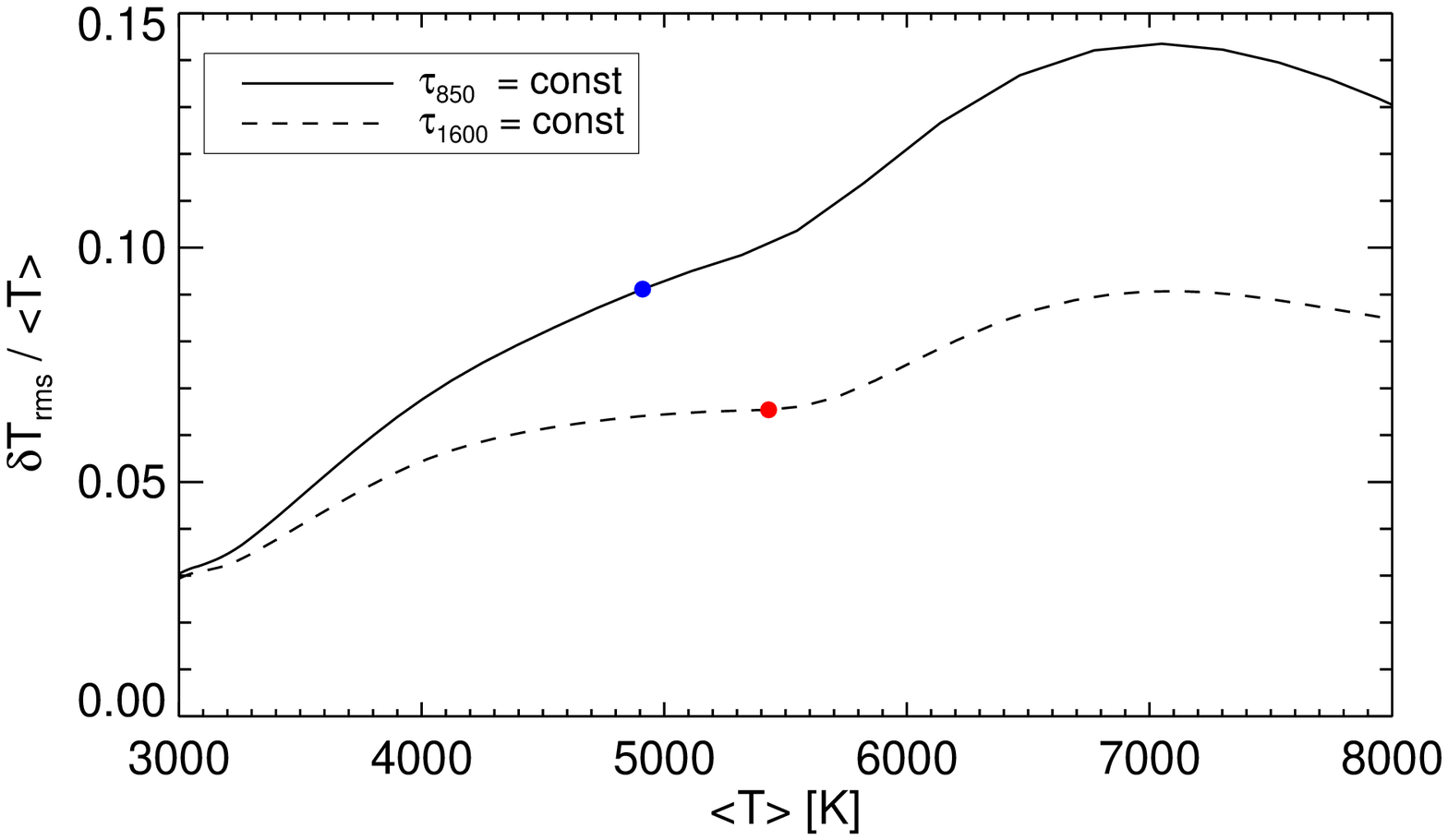}}
\mbox{\includegraphics[bb=32 32 570 380, width=8.9cm]{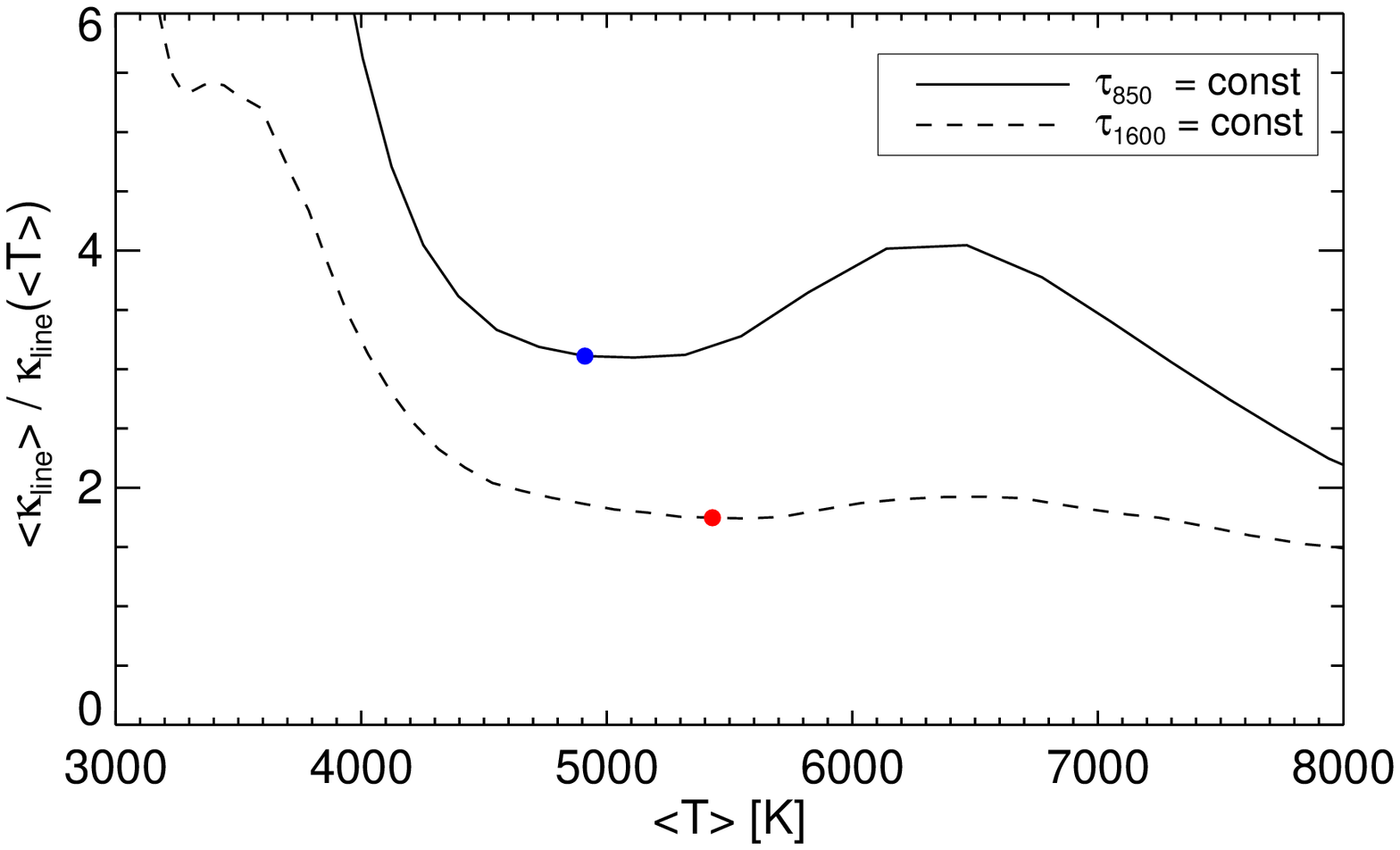}}
\caption{Amplitude of the (relative) temperature fluctuations,
$\delta T_{\rm rms}/\langle T\rangle$ (top), and ratio of average line
opacity to line opacity at mean temperature,
$\mathcal{A}_\ell=\langle\kappa_\ell(T)\rangle/\kappa_\ell(\langle\ T\rangle)$
of a weak (artificial) \ion{Fe}{ii} line with
excitation potential $\chi=10$~eV (bottom) as a function of the mean
temperature $\langle T\rangle$, where angle brackets indicate horizontal
averaging at constant monochromatic optical depth. Solid and dashed curves
show the results of averaging on surfaces of constant $\tau_{850}$ and
constant $\tau_{1600}$, respectively. Filled dots indicate the location of
the center of the line formation regions at
$\lambda\,8500$ and $1600$~nm.}
\label{fig:dTrms}
\end{figure}

Further investigations revealed that the opposite is true. The point is that
we have to distinguish between fluctuations at constant Rosseland optical
depth, $\tauross$, and fluctuations at constant monochromatic optical depth,
$\tauc$, which are relevant in the present context. In fact, the higher
temperature sensitivity of the continuum opacity at
$\lambda$\,$1600$~nm \emph{reduces} the amplitude of the temperature
fluctuations at constant continuum optical depth $\tau_{1600}$ with respect to
the fluctuations at constant $\tau_{850}$, as illustrated in
Fig.\,\ref{fig:dTrms} (top panel). The degree of nonlinearity of
the line opacity fluctuations, as measured by the ratio of average line
opacity to line opacity at mean temperature,
$\mathcal{A}_\ell=\langle\kappa_\ell(T)\rangle/\kappa_\ell(\langle\ T\rangle)$,
is shown in the bottom panel of Fig.\,\ref{fig:dTrms}. Over the whole depth
range, the nonlinearity of the $\kappa_\ell$ fluctuations is higher at
constant $\tau_{850}$ than at constant $\tau_{1600}$. Remarkably,
$\mathcal{A}_\ell$
increases toward lower temperatures, even though the amplitude of the
temperature fluctuations decreases with height. This is because the
temperature sensitivity of $\kappa_\ell$ increases strongly as \ion{Fe}{ii}
becomes a minority species at lower $T$ (cf.\ Fig.\,\ref{fig:numb-dens}).
The fact that $\mathcal{A}_\ell$ is significantly higher for the red line at
$\lambda$\,$850$~nm than for the near-IR line at $\lambda$\,$1600$~nm
explains the wavelength dependence of the abundance corrections found for
case (2), columns (3) and (4).

Comparing cases (2) and (6) for the near-IR line, we see that the abundance
correction essentially vanishes when combining the fluctuations of the
line opacity with the fluctuations of the continuum opacity. We note
that the fluctuations of $\kappa_{\rm c}$ are significantly larger at
$\log \tau_{1600}=0.15$, where the near-IR line forms, than at
$\log \tau_{850}=0.7$, where the red line forms, even though the
temperature fluctuations are lower (see
Fig.\,\ref{fig:dTrms}). This is again a consequence of the
enhanced temperature sensitivity of the continuum opacity
at $\lambda$\,$1600$~nm (Fig.\,\ref{fig:kappa_c}).
It thus happens that the abundance corrections
due to the fluctuations of the continuum opacity and the line opacity,
respectively, are comparable (compare cases 2 and 4). The net result is
a cancelation of the two effects. The total
$\Delta_{\mathrm{3D}-\langle\mathrm{3D}\rangle}$ abundance correction
$\lambda$\,$1600$~nm is therefore small.

\subsection{Saturation effects}
\label{sect:AB_sat}

\begin{figure}[tb]
\centering
\mbox{\includegraphics[bb=32 32 570 380, width=8.9cm]{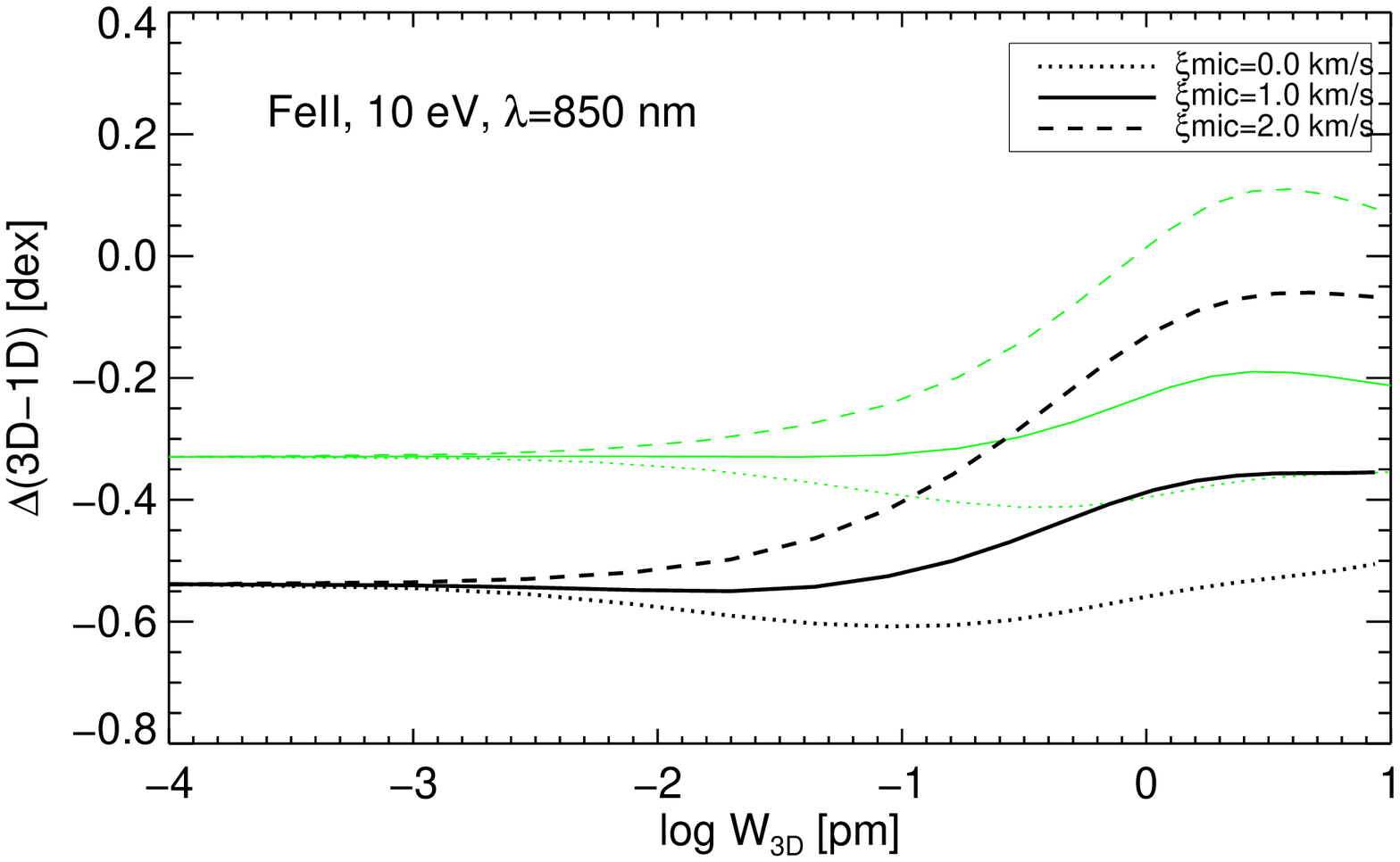}}
\mbox{\includegraphics[bb=32 32 570 380, width=8.9cm]{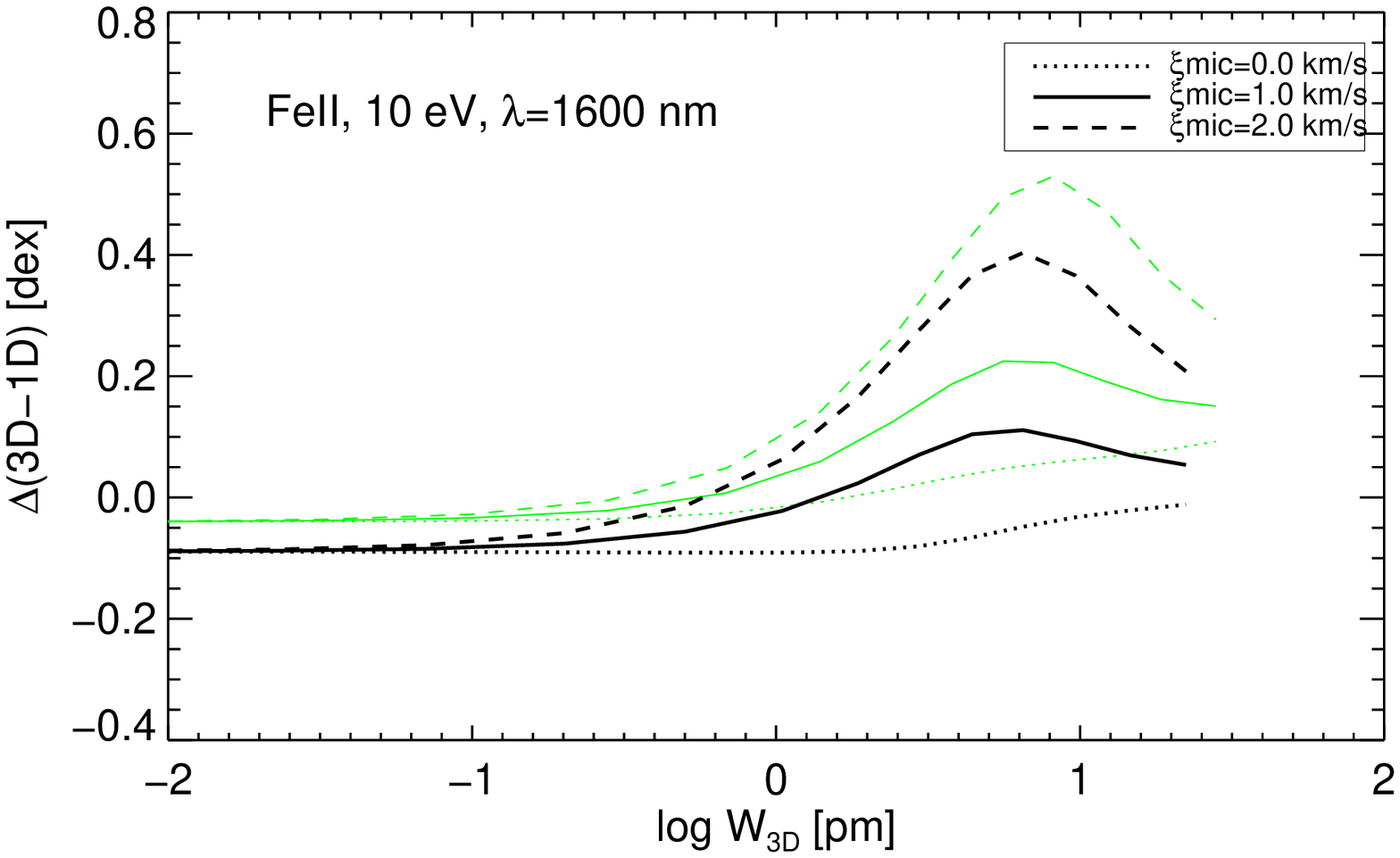}}
\caption{Total 3D abundance correction
$\Delta_{\mathrm{3D}-\mathrm{1D}}$ for the artificial
\ion{Fe}{ii} line with excitation potential $\chi=10$~eV at $\lambda\,850$~nm
(top) and $\lambda\,1600$~nm (bottom) as a function of the equivalent
width obtained from the 3D model. The fainter (green) curves and
the thicker (black) curves refer to the intensity ($\mu=1$) and flux spectrum,
respectively. The abundance corrections have been computed for three different
values of the microturbulence parameter used with the 1D model,
$\xi_{\rm mic}=0.0$ (dotted), $1.0$ (solid), and $2.0$~km/s (dashed lines).
The weak line limit coincides with the horizontal part of the curves at low
$\log W_{\rm 3D}$.}
\label{fig:ac_ew}
\end{figure}

So far we have considered the abundance corrections for the limiting case
of weak, unsaturated lines. In this limit, the abundance corrections are
independent of the equivalent width of the line and of the microturbulence
parameter $\xi_{\rm mic}$ chosen for the spectrum synthesis with the 1D models.
Figure\,\ref{fig:ac_ew} shows how the results change if saturation effects
are fully taken into account, again for the example of the high-excitation
\ion{Fe}{ii} line.

\begin{figure}[tb]
\centering
\mbox{\includegraphics[bb=32 32 570 380, width=8.9cm]{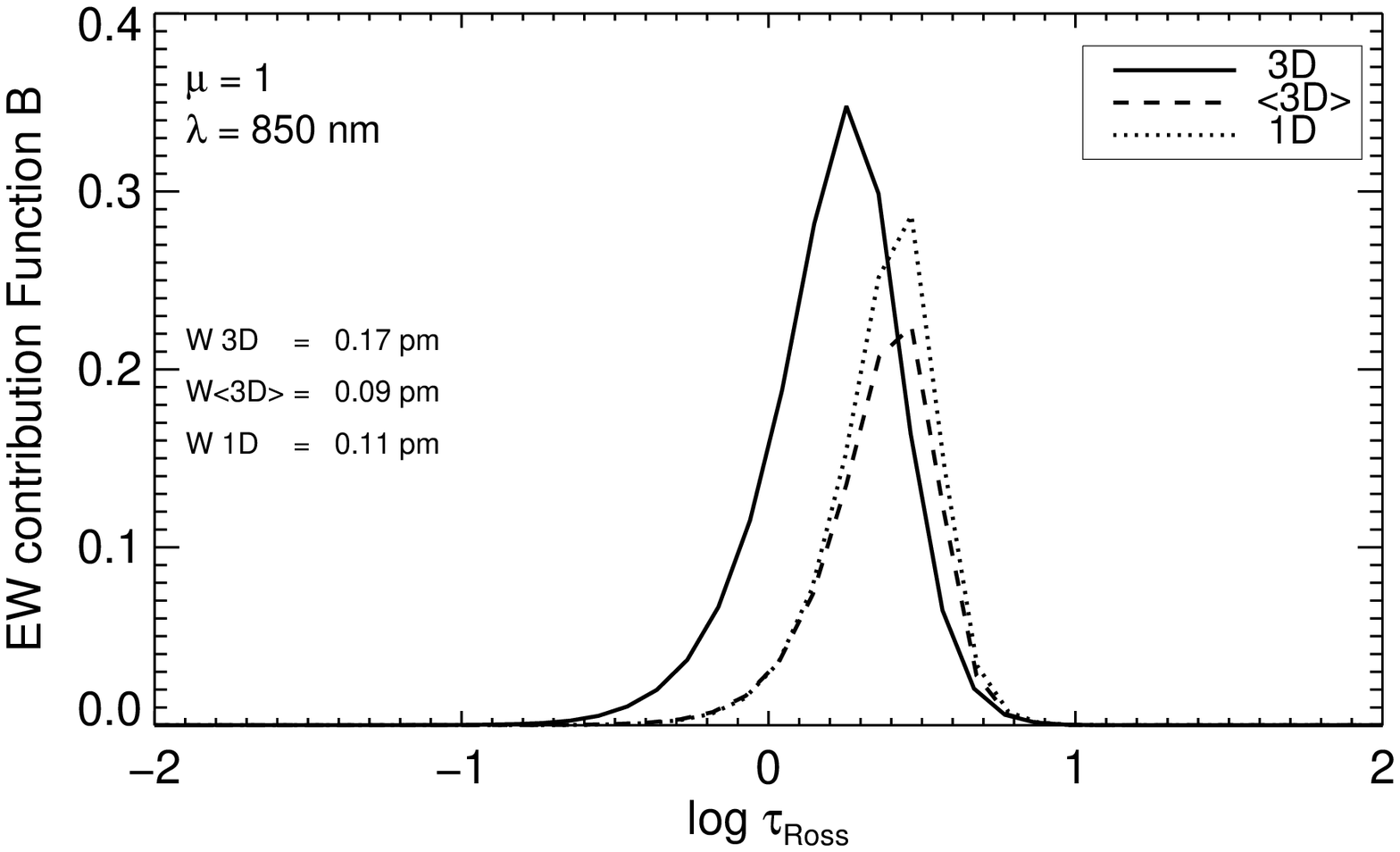}}
\mbox{\includegraphics[bb=32 32 570 380, width=8.9cm]{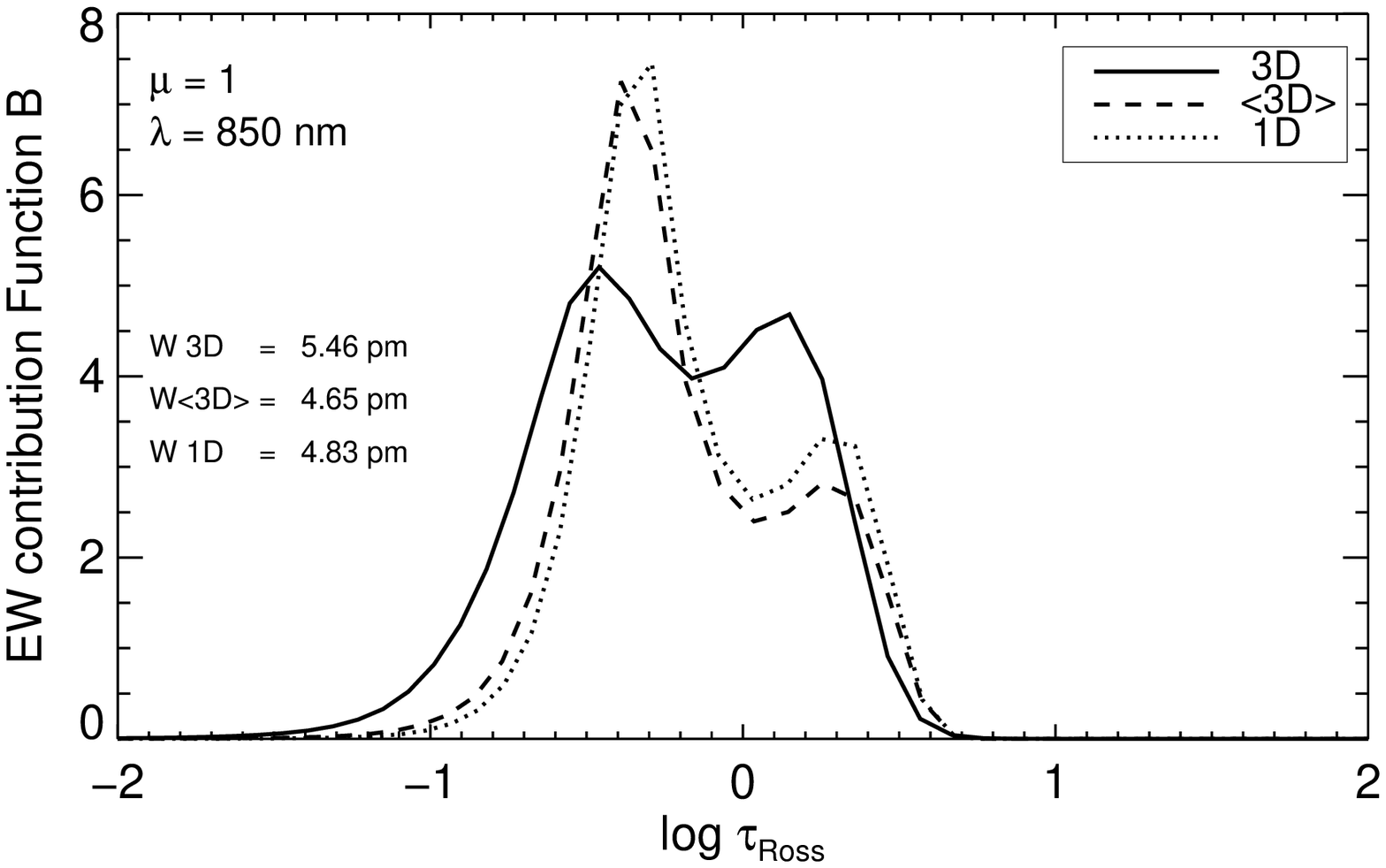}}
\caption{Disk-center ($\mu=1$) equivalent width contribution functions,
$\mathcal{B}(\log \tauc)$, of a weak (top) and strong (bottom)
\ion{Fe}{ii} line with excitation potential $\chi=10$~eV, at
$\lambda\,850$~nm. The contribution functions have been computed for a
single snapshot of the 3D model (solid), the corresponding
$\langle$3D$\rangle$ average model (dashed), and the associated 1D LHD model
(dotted) used in this work. They have been transformed from the monochromatic
to the Rosseland optical depth scale. In all cases, the saturation
factor $exp\{-\taul\}$ is properly taken into account
(see Eq.\,\ref{eqn:B1}).}
\label{fig:cfFeIIsat}
\end{figure}

Obviously, the total 3D abundance correction,
$\Delta_{\mathrm{3D}-\mathrm{1D}}$, depends strongly on both the
assumed value of $\xi_{\rm mic}$ and on the line strength, $W_{\rm 3D}$.
We notice that this
holds even for very weak lines, and conclude that even the weakest lines used
for this study are already partly saturated.
Plotting $\log W_{\rm 3D}$ versus $\log gf$ reveals that the
curve-of-growth is linear, and thus saturation effects can be safely
ignored as long as the equivalent width of the line is below
$W_{\rm 3D}^\ast\approx0.001$~pm at $\lambda\,850$~nm
($W_{\rm 3D}^\ast\approx0.1$~pm  at $\lambda\,1600$~nm). As soon as this
line becomes detectable, it is no longer on the linear part of the curve-of-growth.
This extreme behavior is of course related to the extreme temperature
sensitivity of this high-excitation line, which changes the line-to-continuum
opacity ratio from $\eta \ll 1$ to $\eta \gg 1$ within the line
formation region. This is especially true at $\lambda\,850$~nm, where the
continuum opacity is less temperature dependent (see above). The partial
saturation of weak lines is not a particular property of the 3D model,
but is seen in 1D models, too.

The top panel of Fig.\,\ref{fig:cfFeIIsat} shows the equivalent width
contribution functions $\mathcal{B}(\tauross)$ of the same weak
\ion{Fe}{ii} line ($\chi=10$~eV, $\lambda\,850$~nm) as in
Fig.\,\ref{fig:cfFeII} (top), but now including the saturation
factor $\exp\{-\tau_\ell\}$ (see Eq.\,\ref{eqn:B1}).
Comparison with Fig.\,\ref{fig:cfFeII} (top) demonstrates that including the
saturation factor reduces the equivalent width from $W_{\rm 3D} \approx 0.61$~pm
to $W_{\rm 3D} \approx 0.17$~pm). Moreover, all contribution functions are
shifted to slightly higher layers because of the presence of saturation effects.
The upward shift is more pronounced for the 3D contribution function, because of
the strongly nonlinear fluctuations of the saturation factor
$\exp\{-\tau_\ell\}$. As a result, $\mathcal{B}(\mathrm{3D})$ now becomes
smaller than $\mathcal{B}(\langle\mathrm{3D}\rangle)$ and
$\mathcal{B}(\mathrm{1D})$ in the deepest part of the line-forming region.
Therefore, the ratio of 3D to 1D equivalent width becomes smaller than
in the weak line limit, where $\mathcal{B}(\mathrm{3D}) >
\mathcal{B}(\mathrm{1D})$ over the whole optical depth range. Hence,
the total 3D abundance correction, $\Delta_{\mathrm{3D}-\mathrm{1D}}$, becomes
less negative if saturation is taken into account.

If the line strength is increased even more, the contribution functions become
wider and extend to higher atmospheric layers, as shown in the bottom panel
of Fig.\,\ref{fig:cfFeIIsat}. Recalling that the equivalent width contribution
function is a superposition of the line depression contribution functions
for the individual wavelength positions in the line profile, it seems evident
that the double peak structure is related to the contributions of the line
core (left peak) and of the extended line wings (right peak). Test
calculations confirm this interpretation. Proceeding form the top to
the bottom of the line formation region, the difference
$\mathcal{B}(\mathrm{3D}) - \mathcal{B}(\mathrm{1D})$ changes sign from
positive to negative to positive to negative. Because the contributions of the
different layers to the abundance correction cancel partially, a
straightforward interpretation of the resulting abundance correction becomes
difficult. In principle, a detailed analysis of the line depression
contribution functions at individual wavelengths might lead to further
insights. Noting, however, that the situation becomes even more complicated
when considering flux spectra (involving inclined rays), we have some doubts
that such an investigation is worthwhile.

\section{Molecule concentrations, line opacities, and height of formation}
\label{sect:AC}

The equilibrium number density of diatomic molecules with constituents
$A$ and $B$, $N_{AB}$, is given by the Saha-like relation
\beq
N_{AB} = Q_{AB}(T)\,\frac{N_A}{U_A(T)}\,\frac{N_B}{U_B(T)}\,
\left(\frac{h^2}{2\pi\, m_{AB}\, k T}\right)^{3/2}\,\mathrm{e}^{~D_0/kT}\, ,
\label{eqn:C1}
\eeq
where $N_A$ and $N_B$ are the number densities (per unit volume) of free
neutral atoms (in the ground state) of elements $A$ and $B$, with partition
functions $U_{A}$ and $U_{B}$, respectively; the molecule is characterized
by its mass, $m_{AB}$, its partition function, $Q_{AB}$, and its dissociation
energy $D_0$ \citep[cf.][]{AQ2000}. Defining the number densities per unit
mass as $X_i = N_i/\rho$, where $\rho$ is the mass density, we obtain
\beq
X_{AB} = \rho\,\,Q_{AB}(T)\,\frac{X_A}{U_A(T)}\,\frac{X_B}{U_B(T)}\,
\left(\frac{h^2}{2\pi\, m_{AB}\, k T}\right)^{3/2}\,\mathrm{e}^{~D_0/kT}\, .
\label{eqn:C2}
\eeq

\begin{figure}[tb]
\centering
\mbox{\includegraphics[bb=40 48 575 365,width=8.5cm,clip=true]{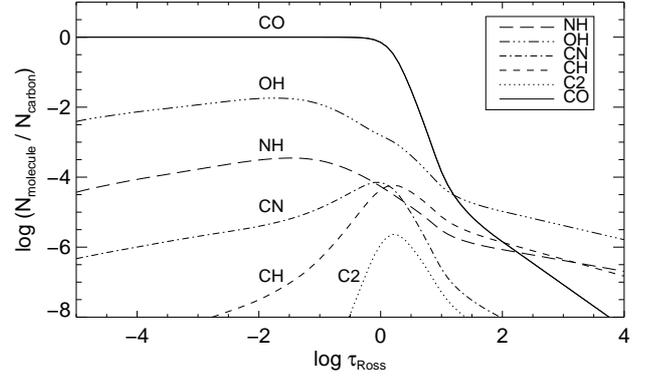}}
\caption{Number density of different molecules, normalized to the total
number density of carbon (sum over all molecules and ionization states)
as a function of the Rosseland optical depth in the 1D LHD model.
\label{fig:frac-molec}}
\end{figure}

\begin{figure}[tb]
\centering
\mbox{\includegraphics[bb=32 32 570 380, width=8.9cm]{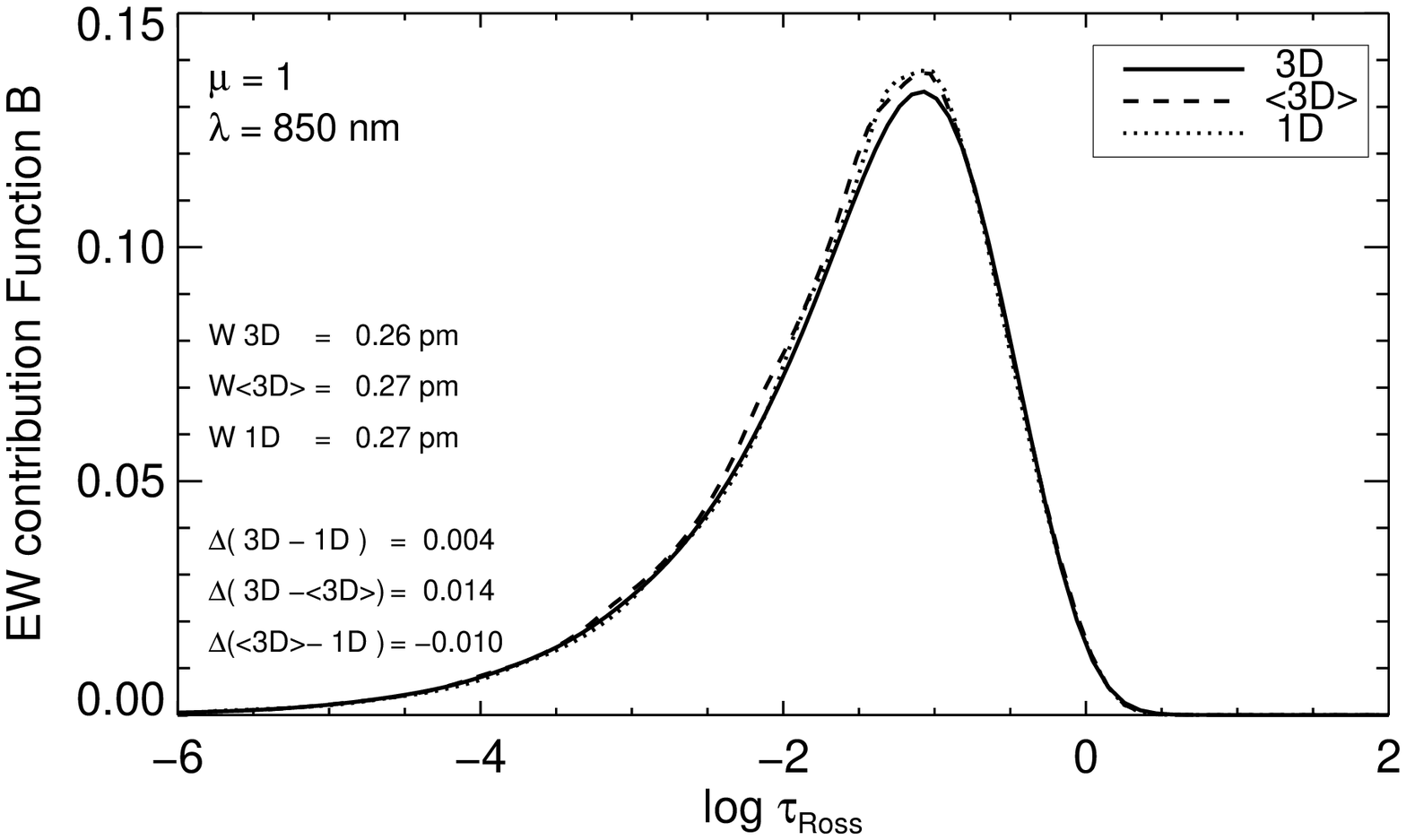}}
\mbox{\includegraphics[bb=32 32 570 380, width=8.9cm]{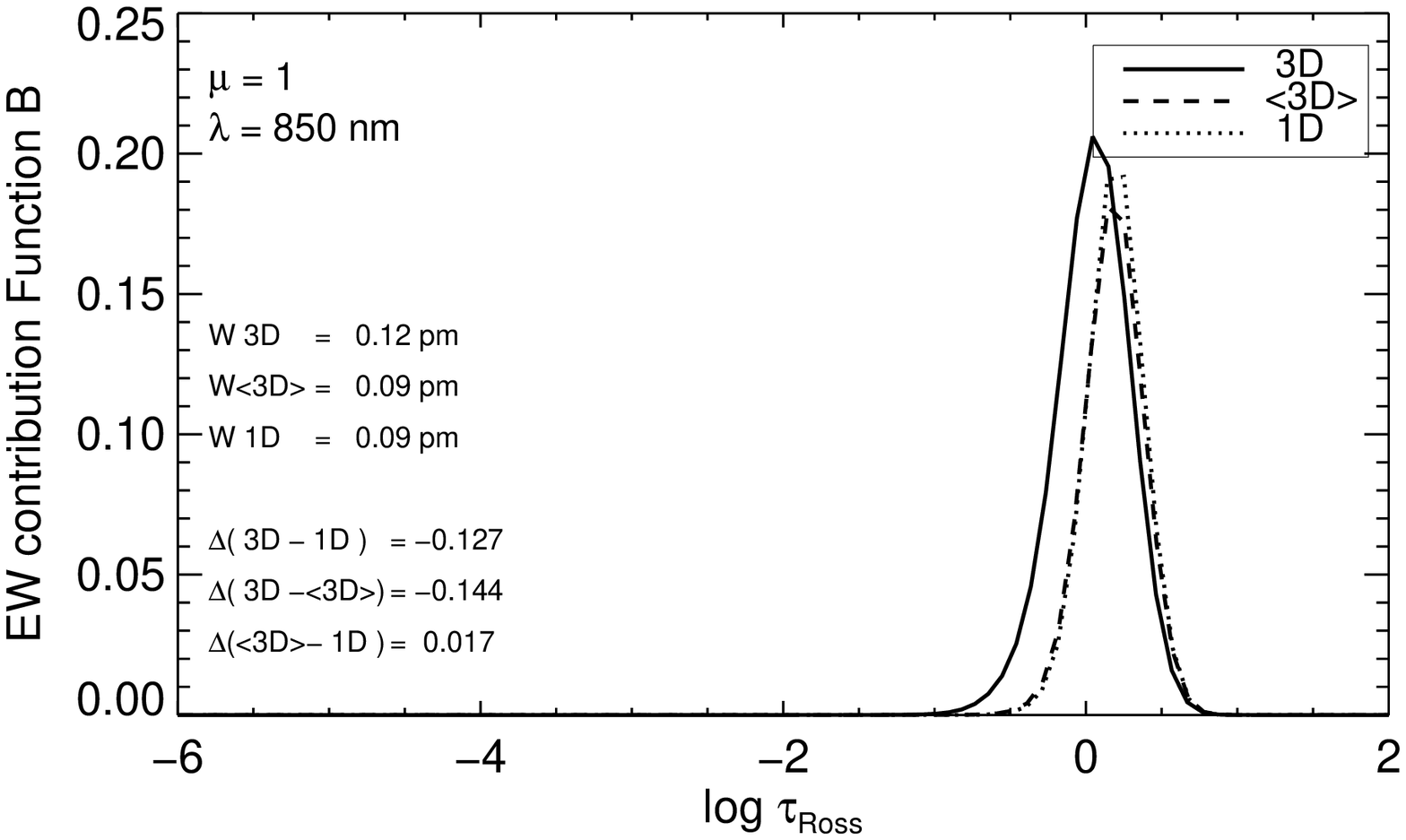}}
\caption{Same as Fig.\,\ref{fig:cfFeI}, but for two weak (artificial)
molecular lines: a CO line with excitation potential $\chi=0$~eV (top)
and a C$_2$ line with $\chi=4$~eV (bottom), both at wavelength
$\lambda\,850$~nm.}
\label{fig:cfmol}
\end{figure}

Figure \ref{fig:frac-molec} shows the number densities of our selection of
diatomic molecules (normalized to the total number of carbon nuclei,
$N_{AB}/\sum N_{\rm C} = X_{AB}/\sum X_{\rm C}$) as a function of Rosseland
optical depth in the 1D LHD model used in this work. In the photosphere ($\log
\tauross < 0$), essentially all carbon is locked up in CO. The decrease
of $X_{\rm OH}$, $X_{\rm NH}$, $X_{\rm CN}$ toward lower optical depths is
a consequence of the density factor $\rho$ in Eq.\,(\ref{eqn:C2}).
The destruction of all molecules beyond $\tauross \approx 1$ is due
to the Boltzmann factor $\exp\{D_0/kT\}$; a higher dissociation energy
corresponds to a steeper drop of the molecule concentration with $T$.

The opacity of a line with lower transition level energy $\chi$,
is proportional to $X_{AB}/Q_{QB}\,\exp\{-\chi/kT\}$, and thus the line
opacity per unit mass can be written as
\begin{eqnarray}
\log \kappa_\ell &=& \log X_{A} + \log X_{B} + \frac{3}{2}
\log \theta + (D_0-\chi)\,\theta \nonumber \\
                &+& \log \rho + \mathrm{const.}\, ,
\label{eqn:C3}
\end{eqnarray}
where $\theta = 5040/T$, and the
temperature dependence of the partition functions $U_A$ and $U_B$
has been ignored; the molecular partition function $Q_{AB}$ cancels out.

If both atoms $A$ and $B$ are majority species (e.g.\ H, N, O for the
conditions in our red giant atmosphere), then $X_A$ and $X_B$ are
constant, and the temperature dependence of the line opacity is given by
\beq\
\frac{\partial{\log \kappa_\ell}}{\partial \theta} = D_0 - \chi +
\frac{3}{2}\,\frac{1}{\theta \ln 10}\, \quad (\rho = \mathrm{const.})
\quad \mathrm{or}
\label{eqn:C4}
\eeq
\beq\
\frac{\partial{\log \kappa_\ell}}{\partial \theta} = D_0 - \chi +
\frac{5}{2}\,\frac{1}{\theta \ln 10}\, \quad (P = \mathrm{const.})\, .
\label{eqn:C5}
\eeq
In general, the temperature dependence of the molecular line opacity is
more complicated, because $X_A$ and/or $X_B$ are more or less strongly
temperature dependent due to ionization and/or formation of different
molecules. In our red giant atmosphere, for example, the concentration
of carbon atoms is controlled by the formation of CO molecules.
This leads to a strong \emph{increase} of $\kappa_\ell$ with temperature
($\partial{\log \kappa_\ell}/{\partial \theta} < 0$) for CH and C$_2$
at $\tauross < 1$, such that these molecules can only form in a narrow
region centered around $\log \, \tauross \approx 0$
(see Fig.\,\ref{fig:cfmol}).

Finally, we point out that the molecular lines form in the same height range as
the lines of neutral atoms and ions. Figure \ref{fig:cfmol} displays the
contribution functions for the most extreme examples. The ground state CO line
(top panel) shows the most extended formation region, centered around $\log
\tauross \approx -1$. The contribution function of this line is almost
identical to that of the ground state \ion{Fe}{i} line shown in
 Fig.\,\ref{fig:cfFeI}. The high-excitation C$_2$ line (bottom panel)
originates from a very narrow formation region located in the deep photosphere
around $\log \tauross \approx 0$. The contribution functions of the other
molecular lines considered in this study lie somewhere in between these two
extremes; the entire formation region of the molecular lines is always
inside the height range covered by our model atmospheres.

\end{appendix}

\end{document}